\documentclass[draft]{agujournal2019}
\usepackage{url} 
\usepackage{lineno}
\usepackage[inline]{trackchanges} 
\usepackage{soul}

\drafttrue

\journalname{JGR: Planets}

\begin{document}

%
%


\title{A stochastic parameterization of non-orographic gravity waves induced mixing for Mars Planetary Climate Model}

%
%




\authors{Jiandong Liu\affil{1,2}, Ehouarn Millour\affil{1}, François Forget\affil{1}, François Lott\affil{1}, Jean-Yves Chaufray\affil{3}}


\affiliation{1}{LMD/IPSL, Sorbonne Université, ENS, Université PSL, École Polytechnique, Institut Polytechnique de Paris, CNRS, Paris, 75005, France.}
\affiliation{2}{Laboratoire de Physique et Chimie de l’Environnement et de l’Espace, CNRS/Université d’Orléans, UMR 7328, Orléans, 45071, France.}
\affiliation{3}{LATMOS, CNRS, Sorbonne Université, Université Versailles St-Quentin, Paris, France}




\correspondingauthor{Jiandong Liu}{jiandong.liu@lmd.ipsl.fr}



\begin{keypoints}
\item Turbulence from non-orographic gravity waves is parameterized in the Mars Planetary Climate Model.
\item Simulations are compatible with observations.
\item Mars has an irregular turbopause depending on the season and local solar time.
\end{keypoints}

%
%

%
%


\begin{abstract}
[ This paper presents a formalism of mixing induced by non-orographic gravity waves (GWs) to integrate with the stochastic GWs scheme in the Mars Planetary Climate Model.
We derive the formalism of GWs and their mixing under the same assumptions, integrating the two schemes within a unified framework. 
Specifically, a surface-to-exosphere parameterization of GW-induced turbulence has been derived in terms of the eddy diffusion coefficient. Simulations show that the coefficient is on the order of 10$^{4}$ to 10$^{9}$ cm$^2$ s$^{-1}$ and a turbopause is at altitudes of 70 to 140 km, varying with seasons. The triggered mixing has minor effects on model temperatures, yet it substantially impacts upper atmospheric abundances. Simulations are consistent with observations from the Mars Climate Sounder and the Neutral Gas and Ion Mass Spectrometer. Mixing enhances the tracer transports in the middle and upper atmosphere, governing the dynamics of these regions. The scheme reveals how non-orographic GW-induced turbulence can regulate upper atmospheric processes, such as tracer escape. ]
\end{abstract}

\section*{Plain Language Summary}
Non-orographic gravity waves are caused by atmospheric motions like convection and jet streams. As these waves rise, they grow and eventually break due to thinning air. This releases energy and creates swirling motions called turbulent eddies at various scales, a phenomenon known as turbulence.
These chaotic swirling motions mix the gases quickly, known as eddy diffusion or mixing. This process is much faster than molecular diffusion, the slow, random spread of molecules from regions of high to low concentration.
This study presents a new way to model eddy diffusion driven by non-orographic gravity waves in Mars’ atmosphere. The method works with existing climate models and uses consistent physics to link wave activity with turbulent mixing.
The strength of mixing, described by the eddy diffusion coefficient, ranges from 10$^4$ to 10$^9$ cm$^2$ s$^{-1}$. The turbopause, the altitude where eddy diffusion and molecular diffusion are equally strong, lies between 70 and 140 km, depending on local time and season.
While this mixing has little effect on temperature, it significantly alters how gases are distributed in the upper atmosphere. The model results match spacecraft observations and show that wave-driven turbulence plays a key role in transporting gases and supporting atmospheric escape on Mars.

%
%

%


%
%
%
%

\section{Introduction}
Martian upper atmospheric dynamics govern the transport and escape of the planet's atmosphere \cite{jakosky2017mars,benna2019global,fedorova2020stormy,stone2020hydrogen,chaffin2021martian,yiugit2021dust}, playing a vital role in understanding the evolution of Mars' habitability \cite{chaffin2017elevated,jakosky2017mars,yiugit2021martian}. Abundances of Martian upper atmospheric neutrals, sampled by the Neutral Gas and Ion Mass Spectrometer (NGIMS) \cite{benna2014datasets,mahaffy2015neutral,england2017maven} and other instruments \cite{creasey2006density,vals2019study,nakagawa2020vertical}, have revealed unexpected density fluctuations \cite{fritts2006mean,mahaffy2015structure,stone2018thermal,liu2019seasonal,li2021horizontal} that cannot be explained by model simulations, even when internal and external forcings are well prescribed  \cite{forget1999improved,gonzalez2009ground,liu2023surface}. 

Generally, orographic gravity waves (GWs) have limited effects on layers above the middle atmosphere \cite{lott1997new,forget1999improved}. Consequently, investigations have focused primarily on forcings from the lower atmosphere and, unsurprisingly, on non-orographic GWs \cite{yiugit2015high, Roeten2022impacts,liu2023surface}. The turbulence generated by non-orographic GWs has been proposed as the cause of the observed density fluctuations  \cite{mcelroy1972stability,mcelroy1976composition,beasley1973wave,rodrigo1990estimates,creasey2006density,fritts2006mean,yoshida2022variations}. 

However, a surface-to-exosphere gravity wave-mixing scheme was not available for a General Circulation Model (GCM), such as the Mars Planetary Climate Model (Mars PCM), for several reasons: (i) There is a lack of full-layer non-orographic gravity wave schemes capable of producing simulations that match observations both below and above the middle atmosphere \cite{Roeten2022impacts,liu2023surface}. Consequently, it is impossible to develop a GW-mixing scheme that is consistent with the gravity wave scheme, as a large portion of the waves become saturated in the thermosphere \cite{beasley1973wave, vals2019study,yiugit2021dust,yiugit2021martian}. (ii) The mechanism of GWs-induced turbulence is not fully understood \cite{liu2021effective}.

The gravity wave Eddy Diffusion Coefficient (EDC) was proposed as a concept analogous to molecular diffusivity to describe the turbulence induced by gravity waves \cite{hodges1967generation,hodges1969eddy,mcelroy1972stability,beasley1973wave,hines1974eddy,lindzen1981turbulence,leovy1982control,holton1982role,holton1988role,strobel1989constraints,shimazaki1989photochemical,barnes1996transport,slipski2018variability,yoshida2022variations}. These turbulence and mixing are proposed to be induced by superadiabatic temperature gradients at or above altitudes where the gravity waves are saturated \cite{hodges1967generation,hodges1969eddy,hines1974eddy,lindzen1981turbulence, holton1982role}. Inspired by this idea, a polynomial relationship between the eddy diffusivity and the vertical wavelength of waves was developed \cite{hodges1969eddy,hines1974eddy}, in which the maximum diffusivity is on the order of 10$^6$ to 10$^8$ cm$^2$ s$^{-1}$ for waves with horizontal wavelengths ranging from 10 to 1000 km. Additionally, an EDC of 10$^6$ to 10$^8$ cm$^2$ s$^{-1}$ has been estimated for the Mars upper atmosphere using this type of relationship \cite{beasley1973wave,strobel1989constraints,barnes1996transport}. All those early attempts are beneficial for 1-D (vertical) monochromatic wave analysis. However, these formalisms may lack consideration of wave saturation and, therefore, are not suitable for a model that includes the upper atmosphere.

\citeA{lindzen1981turbulence} developed a set of equations to describe turbulence caused by wave saturation, assuming that the vertical temperature gradient induced by saturated waves should never surpass the environmental lapse rate. \citeA{holton1982role} and \citeA{weinstock1982nonlinear}  derived a relationship between the pseudo momentum (Eliassen-Palm flux, EP-flux) and the eddy diffusion coefficient, based on the formalism and assumptions of \citeA{lindzen1981turbulence}. Many of the implementations of \citeA{lindzen1981turbulence} and \citeA{holton1982role} have been tested in 3-D GCMs below the thermosphere \cite{lindzen1983turbulence,garcia1985effect,holton1988role,theodore1993solstitial} or in 2-D GCMs that include the thermosphere \cite{imamura2016convective}. Simulations show that the eddy diffusivity influences the vertical transport of chemical tracers in the mesosphere and higher layers \cite{holton1988role,imamura2016convective}. The limitations of \citeA{lindzen1981turbulence} and \citeA{holton1982role} are (i) the saturation conditions are applied too broadly \cite{lindzen1983turbulence,imamura2016convective}, and (ii) the derivations are conducted below the lower thermosphere.

EDC values ranging from 1$\times$10$^4$ to 10$^8$ cm$^2$ s$^{-1}$ have been derived from observations of multiple Martian missions \cite{mcelroy1972stability,nier1976structure,anderson1978mariner,shimazaki1989photochemical,rosenqvist1995reexamination,slipski2018variability,yoshida2022variations}.
These retrievals leverage the life cycle of specific, long-lived atmospheric species (such as CO) \cite{rodrigo1990estimates,yoshida2022variations}, during which fluctuations in the species' abundances must be driven by eddy diffusion \cite{shimazaki1989photochemical,rodrigo1990estimates}. Intensive Ar transport, caused by mixing-circulation interaction, has been observed in the Mars lower atmosphere polar region \cite{sprague2004mars,forget2004alien}, suggesting an EDC of 10$^8$ cm$^2$ s$^{-1}$ during clear-sky seasons and 10$^9$ cm$^2$ s$^{-1}$ during dusty seasons. A power law relationship between eddy diffusivity and the species densities has been found to be realistic for the middle atmosphere \cite{von1980upper,leovy1982control,yoshida2022variations}, implying an EDC of 10$^6$ to 10$^7$ cm$^2$ s$^{-1}$ at altitudes of 70-105 km. 
EDCs estimated from Viking observations are on the order of 10$^7$ to 10$^9$ cm$^2$ s$^{-1}$ at altitudes of 110-170 km \cite{mcelroy1972stability,nier1976structure,mcelroy1976composition}. \citeA{rodrigo1990estimates} estimated an EDC ranging from 10$^7$ to 10$^8$ cm$^2$ s$^{-1}$ for the upper atmosphere below 160 km, with a turbopause at 140 km. \citeA{shimazaki1989photochemical} suggested that an EDC of 5$\times$10$^7$ cm$^2$ s$^{-1}$ is needed to reproduce atomic oxygen observations in the upper atmosphere made by Mariner 6 and 7. The Homopause, inferred from NGIMS-observed N$_2$/$^{40}$Ar ratio \cite{mahaffy2015structure}, shows substantial variations from 75 to 140 km, depending on the season \cite {jakosky2017mars}. Further investigation suggests a co-evolution of the turbopause with the homopause \cite{slipski2018variability}.
All these retrieved turbulence coefficients provide quantitative constraints for a realistic GW turbulence scheme.

The mixing scheme described in this paper builds on the foundation of the surface-to-exosphere gravity wave scheme developed by \citeA{liu2023surface}, which is an improvement and vertical extension of \citeA{lott2012stochastic}. We revisit the seminal formalism proposed by \citeA{lindzen1981turbulence}, \citeA{holton1982role}, and \citeA{weinstock1982nonlinear}, in which turbulence is assumed to be generated by wave saturation. This idea has now been refined as 'turbulence arises from the divergence of wave momentum'. This new concept is then applied to the entire atmosphere using wave-ensembles theory designed by \citeA{lott2012stochastic} and \citeA{lott2013stochastic}. 
In section 2, we review the derivation of the non-orographic GWs formalism using the extended Non-Superadiabatic Principle (NSP). The goal is to provide a theoretical interface for the induced mixing formalism.
In section 3, a mixing scheme is designed by using the extended NSP and the theoretical interface. A comprehensive formula for the eddy diffusion coefficient is derived from both theoretical and empirical analyses. Section 4 presents numeric simulations using the Mars Planetary Climate Model. The simulated results are compared with observations from both MCS and NGIMS. Summaries and future perspectives are outlined in section 5.

\section{Non-orographic Gravity Waves}
 The original design of the non-orographic gravity wave scheme described in this paper was developed by \citeA{lott2012stochastic}, in which the waves are represented using stochastic wave ensembles \cite{lott2013stochastic}. This elegant design accumulates wave momentum over a few model time steps and then employs a first-order Auto-Regressive algorithm (AR-1) to transfer the momentum to the mean zonal flow. The scheme performs well for the LMDZ.EARTH model, which excludes the upper atmosphere \cite{de2014intermittency}, successfully reproducing the well-known Quasi-Biennial Oscillation (QBO) in the Earth's mesosphere \cite{lott2013stochastic}. The design was adapted and refined by \citeA{liu2023surface} to operate within the Mars Planetary Climate Model (Mars PCM), a surface-to-exosphere GCM \cite{gonzalez2009ground}. The adapted scheme reproduces temperature structures below 100 km consistent with MCS observations and simulates upper atmospheric abundances compatible with NGIMS measurements \cite{liu2023surface}.

We consider that at each model (physical) step $ t$ (of duration $\delta t $), the wave-induced vertical velocity $w'$ is represented as a weighted sum of $M$ monochromatic waves' amplitudes $\hat{w}_j$ : $w' = \sum_{j=1}^{M} C_{j}\hat{w}_{j}$, where the weights satisfy $\sum_{j=1}^{M} C_{j}^{2} =1$. Following \cite{lott2012stochastic}, we use ${M=8}$ waves. 
To evaluate the wave amplitude $\hat{w}_{j} $, we randomly assign the horizontal wavenumber $k_{j}$ and $l_{j}$, phase speed $c_j$, orientation $\vartheta \in [0,2\pi]$, and initial momentum flux at a specific reference (launch) altitude $ z_{r} $. The vertical propagation of $\hat{w}_{j} $ from one model level $ z_{ll} $ to the next $ z_{ll+1} $ is then treated by a Wentzell-Kramers-Brillouin (WKB) approximation. 

In this section, we will present the detailed formalism of \citeA{liu2023surface} to provide a common framework for both non-orographic GWs and wave-induced turbulence. 

\subsection{ Eliassen-Palm Flux}
Rather than treating the waves directly, their effects are accounted for through their pseudo momentum, specifically the Eliassen-Palm (EP) flux \cite{fritts1984gravity,yiugit2008parameterization,lott2012stochastic,lott2013stochastic}. When rotation is neglected, the EP-flux is assumed to be a constant at its launch altitude \cite{fritts2003gravity}. For $j$th monochromatic wave at the launching (reference) level $z_r$, where the waves are generated, the EP-flux is given by,
\begin{equation}
\vec{\textit{E}}_j^{z_r}(k_j,l_j,\omega_j)=\Re \bigg\lbrace \rho_r \frac{\hat{u} \hat{w}^*}{2} \bigg\rbrace
\label{EPF}
\end{equation}
where $\rho_r$ is the atmospheric mass density at the launching (reference) altitude; $\hat{u}$ and $\hat{w}$ are zonal and vertical wind amplitudes; $\Re$ denotes the real part; $k_j$,$l_j$ horizontal wavenumbers, and $\vec{k}=(k_j,l_j)$; $\omega_j=\vec{k_j}c_j$ is the absolute frequency of the wave, with $c_j$ representing the phase speed. 
In the Mars PCM, the maximum truncation of $\vec{E}_j^{z_r}$ equals 5$\times$10$^{-4}$ kg m$^{-1}$ s$^{-2}$  \cite{liu2023surface}.

The zonal wind amplitude $\hat{u}$ depends on the vertical wind amplitude $\hat{w}$ through waves' vertical wave number $m$, as described by the "polarized material equation" \cite{fritts1984gravity,fritts2003gravity},
\begin{equation}
\hat{u}=-\frac{m}{|\vec{k}|} \hat{w} \frac{\vec{k}}{|\vec{k}|}
\label{UWM}
\end{equation}
The EP-flux can be rewritten as a function excluding $\hat{u}$  by inserting (\ref{UWM}) into (\ref{EPF}), which reads,
\begin{equation}
\vec{E}_j^{z_r}(k_j,l_j,\omega_j)=\Re \bigg\lbrace \rho_r \frac{\hat{u} \hat{w}^*}{2} \bigg\rbrace = - \rho_r \frac{\vec{k}}{2 \vert \vec{k} \vert^2} m_j(z_r) \vert \hat{w}_{j}(z_r) \vert^2
\label{fullEPF}
\end{equation}
Equation (\ref{fullEPF}) provides an expression for the EP-flux that includes solely the wave amplitude $\hat{w}_j$ (For convenience, the index $j$ is dropped from $m$, $k$, $l$, and $c$ ). This implies that the EP-flux of a specific wave $j$ depends solely on $\hat{w}_j$. Consequently, we will search for $\hat{w}_j$ from an appropriate wave equation within a hydrostatic atmosphere in the following sections.

\subsection{Taylor-Goldstein Equation}
The Taylor-Goldstein Equation (TGE) describes the evolution of the amplitude of small perturbations (i.e., $\hat{w}_j$ in this paper) in stably stratified, incompressible shear flows \cite{fritts1984gravity,lindzen1981turbulence}. 
It has been widely used as the governing equation of the gravity wave theory \cite{yiugit2008parameterization,lott2012stochastic}, with its derivation detailed in references such as \citeA{nappo2013introduction}. 

The key to evaluating the life cycle of a specific monochromatic wave lies in tracking the behavior of $\hat{w}_j$, which is governed by the wave's pseudo momentum as expressed in equation (\ref{fullEPF}).
Assuming a gravity wave with absolute (intrinsic) frequency $\omega_j \gg f$ (the Coriolis frequency), the vertical velocity (perturbation) $w'_j$ of the $j$th harmonic satisfies,
\begin{equation}
w'_{j} = \Re \lbrace \hat{w}_{j}(z) e^{z/2H}e^{i(k_{j}x+l_{j}y-\omega_{j} t)} \rbrace 
\label{vwe}
\end{equation}
in which case the amplitude $\hat{w}_j$ satisfies the following Taylor-Goldstein Equation (TGE),
\begin{equation}
\frac{\partial^2 \hat{w}(z)}{\partial z^2}+ \bigg(\underbrace{\frac{\vert \vec{ k}  \vert^2 N^2}{\Omega^2}+ \frac{\vec{k}(\vec{u}_{zz}+\vec{u}_z /H)}{\Omega} -\frac{1}{4H^2}}_{Q(z)} \bigg)\hat{w}(z) = 0
\label{fullTGE}
\end{equation}
 where $ N^2 =g/T\Gamma $ represents the square of Brunt-Väisälä frequency; $\Gamma=\partial T/\partial z+g/c_{p}$ is the environmental lapse rate; $g$ is the gravity acceleration, $T$ the temperature, and $c_{p}(T)$ the specific heat capacity.
 The vertical coordinate $ z $ is the pseudo-altitude, defined as $ z=H\ln(P_{r}/P) $, where $ P $ is the pressure, $ P_{r} $ the pressure at the wave reference altitude, and $ H $ the atmospheric scale height. $\vec{u}_{zz}$ is $\partial^2 \vec{u}/ \partial z^2$ and $\vec{u}_{z}$ is $\partial \vec{u}/ \partial z$;
 $\Omega=\vec{k}(\vec{k}c/|\vec{k}|-\vec{\bar{u}})$ is the wave's intrinsic frequency with Doppler shift; $\bar{u}$ the background winds. 

In most circumstances, the last two terms of $Q(z)$ in equation (\ref{fullTGE}) are negligible, i.e., $Q(z)\approx \vert \vec{k}\vert ^2 N^2 / \Omega^2  $.
Therefore, equation (\ref{fullTGE}) simplifies to,
\begin{equation}
\frac{\partial^2 \hat{w}(z)}{\partial z^2}+ m^2_r \hat{w}(z) = 0
\label{simpTGE}
\end{equation}
We define this approximated $Q(z)$ as the square of the (real) vertical wavenumber,
\begin{equation}
m^2_r= \frac{\vert \vec{ k}  \vert^2 N^2}{\Omega^2}
\label{vwn2}
\end{equation}
We label the vertical wave number as $m_r$ instead of $m$, where $m_r=\Re \lbrace m \rbrace$. This distinction is made to differentiate from $m_i=\Im \lbrace m \rbrace$, the imaginary part of $m$, which is induced by the turbulence of the wave. The $m_i$ will be detailed in section 2.4.2. Note that $m_r$ changes slowly with altitude $z$.

All subsequent analysis is based on the wave equation (\ref{simpTGE}). Once the solution $\hat{w}$ of this wave equation is obtained, the wave's momentum flux is calculated using equation (\ref{fullEPF}). The form of $\hat{w}$ depends critically on the sign of the square of the Brunt-Väisälä frequency, $ N^2$, as seen in equations (\ref{simpTGE}) and (\ref{vwn2}); note that $|\vec{k}|^2$ and $\Omega^2$ are both positive. 

\subsection{Wentzell-Kramers-Brillouin Approximation}
If $m^2_r$ were constant, (\ref{simpTGE}) would be trivial to solve. However, since $m^2_r$ varies slowly with $z$, the Wentzell-Kramers-Brillouin (WKB) approximation applies \cite{lindzen1981turbulence,holton1982role}, 
\begin{equation}
\hat{w}(z) \approx A(z) |m_r| ^{-1/2} exp \bigg ( i \int^z_0 m d \zeta \bigg )
\label{WKB}
\end{equation}
where $A(z)$ is an altitude-dependent constant for $\zeta \in$ [surface, top]; again, we define $m=m_r+im_i$ (with $i^2=-1$) but neglect $\Im(m)$ (its physics and derivation follow in sections 2.4.2). The real component $\Re(m)$ is obtained from (\ref{vwn2}) as,
\begin{equation}
m_r=-\frac{N|\vec{k}|}{\Omega}.
\label{mreal}
\end{equation}
Note that $N<0$, if $z>z_r$; thus $m_r>0$ in (\ref{mreal}) for $z>z_r$. The real part of the vertical wavenumber is sometimes represented as $m_r=N/(\bar{u}-c)$ for convenience since $\Omega=\vec{k}(\vec{k}c/|\vec{k}|-\vec{\bar{u}})$.

The WKB approximation can also be expressed recursively in altitude to eliminate the constant $A(z)$. Applying (\ref{WKB}) to adjacent layers $z_{ll}$ and $z_{ll+1}$ and taking their ratios yields,
\begin{equation}
 \hat{w}_{j}(z_{ll+1}) =\hat{w}_{j}(z_{ll}) \sqrt{\frac{m_r(z_{ll})}{m_r(z_{ll+1})}} exp \bigg ( i \int^{z_{ll+1}}_{z_{ll}} m^{ave} d \zeta \bigg )
\label{deuxWKB}
\end{equation}
where $m^{ave}=(m(z_{ll})+m(z_{ll+1}))/2$ and $A(z_{ll}) \approx A(z_{ll+1})$. Both (\ref{WKB}) and (\ref{deuxWKB}) represent Wentzel-Kramers-Brillouin (WKB) approximations of (\ref{simpTGE}) for slowly varying $m^2_r$. Notably, (\ref{deuxWKB}) omits $A(z)$, enhancing its practical utility.

We now have nearly exact solutions for $\hat{w}_j$ through (\ref{WKB}) and (\ref{deuxWKB}), but two obstacles prevent their application to the EP-flux (\ref{fullEPF}): (i) the need to determine $m_i$ in WKB approximations, and (ii) the real atmosphere effects (critical layer, saturation, and kinematical viscosity) modify the vertical evolution of $\hat{w}_j$. Both challenges are addressed by incorporating thermodynamic constraints during wave saturation \cite{lindzen1981turbulence} and accounting for kinematic viscosity \cite{lott2012stochastic}.

\subsection{Saturation, Critical Layer and Atmosphere Viscosity}
Three factors govern the vertical evolution of $\hat{w}_j$: (i) wave saturation ( and induced turbulence damping), where EP-flux peaks and releases (most of) momentum to the mean zonal flow as atmospheric density decreases exponentially \cite{lindzen1981turbulence,lott2012stochastic}; 
(ii) critical layers that nullify wave momentum at $\Omega$ phase transitions \cite{lott2012stochastic,lott2013stochastic}; and (iii) kinematic viscosity, particularly dominant in the upper atmosphere, which damps wave amplitude \cite{yiugit2008parameterization,liu2023surface}. These mechanisms are well-documented observationally \citeA{yiugit2015high,vals2019study,liu2019seasonal}.

\subsubsection{Wave Thermodynamics}
As pointed out by \citeA{hodges1967generation} and \citeA{hodges1969eddy}, turbulence arises from localized superadiabatic regions (convective instabilities) triggered by gravity waves. \citeA{lindzen1981turbulence} noted that wave saturation and the resulting turbulence limit vertical temperature gradients to the adiabatic lapse rate $\Gamma$. This implies that turbulence suppresses the growth of superadiabatic regions \cite{hodges1969eddy}, maintaining vertical temperature gradients near the adiabatic lapse rate $\Gamma$ \cite{fritts1984gravity}. This also suggests that a form of thermodynamic equilibrium is established during wave momentum release in the affected region.

For a monochromatic harmonic with vertical velocity $w'_j$ given by equation (\ref{vwe}), the associated temperature amplitude $\delta \hat{T}$ resulting from the wave's work on the atmosphere can be derived from the thermodynamic energy equation in its polarized form,
\begin{equation}
    i\vec{k_j}(c-\bar{u})\delta \hat{T} = \hat{w}_j e^{z/2H} \Gamma, \quad z\in[surfce,top]
    \label{thee}
\end{equation}
where $\Gamma=dT/dz+g/c_{p}$ is the environmental lapse rate. Notice $\Omega=\vec{k}(c-\bar{u})$, we can write it as $ i \Omega \delta \hat{T} = \hat{w}_j e^{z/2H} \Gamma$ . Note that the index $j$ for phase speed and wavenumbers has been omitted for convenience.

From the thermodynamic energy equation (\ref{thee}), one obtains,
\begin{equation}
     \delta \hat{T} = \frac{\Gamma \hat{w}_j e^{z/2H}} {i \Omega}
     \label{deltaT01}
\end{equation}
Furthermore, we replace the vertical velocity amplitude $\hat{w}_j$ in (\ref{deltaT01}) with its WKB approximation from (\ref{WKB}), using $m_r = -N |\vec{k}| / \Omega$.
Equation (\ref{deltaT01}) then becomes,
\begin{equation}
   \delta \hat{T} 
     =\frac{-i\Gamma  A |m_r| ^{-1/2} e^{i \int^z_0 m d \zeta} e^{z/2H}  } { \Omega}  
     \label{incTFULL}
\end{equation}
The $\delta \hat{T}$ in (\ref{incTFULL}) depends on the stochastic parameters of a monochromatic wave and the background atmosphere, including the vertical wavenumber $m$, horizontal wavenumber $\vec{k}$, wave intrinsic frequency with Doppler shift $\Omega$, environmental lapse rate $\Gamma$, and atmospheric density $\rho\propto e^{z/2H}$. The vertical evolution of $\delta \hat{T}$ is primarily governed by $m_r$, $\exp{\int^z_0 m d\zeta}$, and $\exp{z/2H}$, while other terms are constants.

Let $L= -i\Gamma  A |m_r| ^{-1/2} e^{i \int^z_0 m d \zeta} e^{z/2H}$, and thus 
$\delta \hat{T} = L/\Omega$. 
By differentiating equation (\ref{incTFULL}) with respect to $z$, the vertical temperature gradient induced by the wave's vertical velocity follows,
\begin{equation}
      \frac{d \delta \hat{T}}{dz} = \frac{L}{\Omega^2} \bigg [ \Omega \bigg ( im  +\frac{1}{2H}   \bigg)-\frac{\partial \Omega}{\partial z}    \bigg]
     \label{tempgc}
\end{equation}
Considering $m=m_r+im_i$ and the vertical wavenumber $m_r$ in (\ref{mreal}), the real value of the temperature gradient (\ref{tempgc}) reads,
\begin{eqnarray}
\Re \bigg \lbrace \frac{d \delta \hat{T}}{dz} \bigg \rbrace &=& \frac{\Gamma  A |m_r| ^{-1/2} e^{- \int^z_0 m_i d \zeta} e^{z/2H}  m_r}{\Omega}   \nonumber \\
&=&-\frac{\Gamma  A |m_r| ^{3/2} e^{- \int^z_0 m_i d \zeta} e^{z/2H}  }{N |\vec{k}|}   \nonumber \\
&=&-\frac{\Gamma  A |m_r| ^{3/2} e^{ \int^z_0 (\frac{1}{2H}-m_i )d \zeta}  }{N |\vec{k}|}, \quad z\in [surface,top]
     \label{rtempg}
\end{eqnarray}
$A\Gamma/N|\vec{k}|$ varies slowly for a given harmonic. 
The terms $e^{z/2H}$, $\vert m_r \vert^{3/2}$, and $\exp{\int^z_0 m_i d \zeta}$ dominate the growth of $ \vert \Re (\frac{d\delta \hat{T}}{dz}) \vert$ as shown in (\ref{rtempg}).
$\vert m_r \vert^{3/2}$ and $\exp{-\int^z_0 m_id \zeta}$ are induced by the wave-mean flow interaction.
However, equation (\ref{rtempg}) cannot be evaluated directly due to the unknown parameters $m_i$ and $A$, requiring additional constraints for the solution.

\subsubsection{Non-Superadiabatic Principle and $m_{i,s}$}
\citeA{hodges1967generation,hodges1969eddy} were the first to explain that turbulence observed in the upper atmosphere can result from localized superadiabatic regions ($\vert \Re (\frac{d\delta \hat{T}}{dz}) \vert > \Gamma$) induced by wave saturation. This turbulence, once triggered, suppresses further growth of these regions. This concept was implemented in practice by \citeA{lindzen1981turbulence} and \citeA{holton1982role}, assuming that the superadiabatic region within the thin layers around the wave saturation altitude is suppressed completely by the induced turbulence. We refer to this assumption as the Non-Superadiabatic Principle (NSP).

While the NSP provides a valuable framework by connecting turbulence to wave saturation \cite{hodges1967generation,hodges1969eddy,lindzen1981turbulence}, or applying it thereafter \cite{holton1982role,weinstock1982nonlinear}, there are a couple of aspects that might benefit from further refinement: (i) it could lead to an overestimation of turbulence above the saturation altitude, and (ii) it doesn’t fully capture the turbulence below, which is observed in practice.

We reinterpret the NSP without altering the formalism of \citeA{lindzen1981turbulence} and \citeA{holton1982role}, but limit its scope. This includes four key points:

(a) Turbulence arises from the unstable region caused by the wave's energy release to the mean flow (observed wave-like temperature fluctuations), meaning eddy diffusivity is proportional to the divergence of wave momentum ($D_{eddy}\propto \nabla(\vec{E}_j)$).

(b) Therefore, turbulence occurs both below and above the saturation altitude (symmetry).

(c) Turbulence limits or eliminates the growth of the influenced region. To prevent excessive temperature gradient growth with altitude during momentum release, the term $\vert m_r \vert^{3/2}$ must balance the exponential term $exp[\int^z_0 (1/2H -m_i)d \zeta]$ in \ref{rtempg}.

(d) At the saturation altitude $z_b$, the releasing of wave’s momentum and the induced turbulence reach their maxima, as the divergence of the EP-flux peaks at $z_b$ \cite{fritts1984gravity,yiugit2008parameterization,lott2012stochastic,liu2023surface}.

Considering the vertical wavenumber definition of (\ref{mreal}) and applying the NSP-(c), one has,
\begin{eqnarray}
    \frac{1}{\vert m_r \vert^{3/2}} \frac{d \vert m_r \vert^{3/2}}{dz} &=& \frac{3}{2} \frac{1}{m_r} \frac{d m_r}{dz} \nonumber \\
    &=& \frac{3}{2} \frac{\Omega}{N |\vec{k}|} \frac{N'|\vec{k}|\Omega-N|\vec{k}| \Omega'}{\Omega^2} \nonumber \\
    &=&\frac{3}{2} \bigg\vert\frac{1}{N} \frac{\partial N}{dz} - \frac{1}{\Omega} \frac{\partial \Omega}{dz} \bigg\vert, \quad z\in [surface, top]
    \label{mexp}
\end{eqnarray}
The left side of (\ref{mexp}) is designed to form the term $\ln{\vert m_r \vert^{3/2}}$ when we integrate both sides of the equation. 
Solving equation (\ref{mexp}) gives the following results:
\begin{equation}
    \vert m_r \vert^{3/2} = \exp \bigg[\frac{3}{2} \int^z_0 \bigg\vert\frac{1}{N} \frac{\partial N}{d\zeta} - \frac{1}{\Omega} \frac{\partial \Omega}{d\zeta} \bigg\vert d\zeta \bigg], \quad \zeta\in [surface,top]
    \label{mexps}
\end{equation}
The expression in (\ref{mexps}) takes account of the vertical variation of the Brunt-Väisälä frequency $N$, which has been neglected by \citeA{lindzen1981turbulence} and \citeA{holton1982role}. The factor cannot be neglected in the upper atmosphere, especially in the region where the saturation is taking place \cite{hodges1967generation}.

To balance the terms in the temperature gradient of (\ref{rtempg}) as suggested by NSP-(c), we substitute the $\vert m_r \vert^{3/2}$ expression from (\ref{mexps}) into (\ref{rtempg}), yielding,
\begin{equation}
    \Re \bigg \lbrace \frac{d \delta \hat{T}}{dz} \bigg \rbrace =  -\frac{\Gamma A}{ N |\vec {k}|} \exp \bigg \lbrace \int^z_0 \bigg [ \frac{1}{2H} + \frac{3}{2} \bigg\vert\frac{1}{N} \frac{\partial N}{\partial \zeta} - \frac{ 1}{\Omega} \frac{\partial \Omega}{\partial \zeta}\bigg\vert -m_i  \bigg] d\zeta \bigg \rbrace, \quad \zeta\in [surface,top]
    \label{nonsp}
\end{equation}
The $j$th harmonic gets saturated when the (\ref{nonsp}) hits its extrema as implied by NSP-(d). We label the $m_i$ at saturation altitude $z_b$ as $m_{i,s}$, i.e., saturated vertical wavenumber of turbulence,
\begin{equation}
    \frac{1}{2H} + \frac{3}{2} \bigg\vert\frac{1}{N} \frac{\partial N}{\partial z} - \frac{ 1}{\Omega} \frac{\partial \Omega}{\partial z}\bigg\vert = m_{i,s}, \quad   z = z_b, \quad \forall j\in[1,M]
    \label{imagem}
\end{equation}

Note that equation (\ref{imagem}) applies only at the saturation altitude $z_b$ for a given monochromatic wave. \citeA{lindzen1981turbulence} and \citeA{holton1982role} extended (\ref{imagem}) to all atmospheric layers above $z_b$, which is a key difference from our approach. 

\subsubsection{Saturation of The Wave}
As stated in equation (\ref{imagem}) and NSP-(d), the temperature gradient induced by the wave reaches its extrema at the breaking (saturation) altitude $z_b$. According to NSP-(c), the turbulence triggered by the superadiabatic atmosphere prevents the region from expanding by activating the imaginary part of the vertical wavenumber $m_i$ in (\ref{rtempg}), thereby balancing the exponential growth of $ \vert \Re \lbrace \frac{ d \delta \hat{T}}{dz} \rbrace \vert$. As a result, $ \vert \Re \lbrace \frac{ d \delta \hat{T}}{dz} \rbrace \vert$ should not deviate significantly from the environmental lapse rate $\Gamma$ due to the induced turbulence. While \citeA{lindzen1981turbulence} proposed that $\vert \Re \lbrace \frac{ d \delta \hat{T}}{dz} \rbrace \vert = \Gamma$ at and above $z_b$, we argue that this holds only at $z_b$ for a monochromatic wave, as suggested by NSP-(d), providing a constraint to derive the values of $A$ and $z_b$ in equation (\ref{WKB}).
Considering the absolute value of the temperature gradient from (\ref{rtempg}), we get,
\begin{eqnarray}
\bigg \vert \frac{\Gamma  A |m_r| ^{3/2} e^{- \int^{z_b}_{0} m_i d \zeta} e^{z_b/2H}  }{N |\vec{k}|} \bigg \vert &=&\Gamma, \quad z=z_b \nonumber \\
\frac{\Gamma^2  A^2 m_r^{3} e^{- 2\int^{z_b}_{0} m_i d \zeta} e^{z_b/H}  }{N^2 |\vec{k}|^2} &=&\Gamma^2, \quad z=z_b  \nonumber \\
 A^2 m_r^{-1} \exp \bigg(-2 \int^{z_b}_{0} m_i d \zeta \bigg)    &=&m_r^{-4} N^2 |\vec{k}|^2 \exp \bigg(-\int^{z_b}_{0} \frac{1}{H} d \zeta \bigg), \qquad   z=z_b
 \label{ampliA}
\end{eqnarray}

Taking into account (\ref{ampliA}) and the WKB approximation in (\ref{WKB}), the square of the vertical velocity amplitude $\hat{w}_j$ at the saturation altitude $z_b$ is given by,
\begin{eqnarray}
|\hat{w}_j(z_b)|^2 &\approx& A^2 m_r^{-1} \exp \bigg ( - 2  \int^{z_b}_0 m_i d \zeta \bigg )  \nonumber\\
               &=& m_r^{-4} N^2 |\vec{k}|^2 \exp \bigg(-\int^{z_b}_{0} \frac{1}{H} d \zeta \bigg) \nonumber \\
               &=& \frac{\Omega^4}{N^2 |\vec{k}|^2} \exp \bigg(-\int^{z_b}_{0} \frac{1}{H} d \zeta \bigg) 
               \label{ws2}
\end{eqnarray}
The square root of (\ref{ws2}) represents the saturation (breaking) amplitude $\hat{w}_{j,s}$ of a monochromatic wave, initially proposed by \citeA{lindzen1981turbulence}. However, equation (\ref{ws2}) assumes the condition of (\ref{ampliA}), which enforces a strict non-superadiabatic state around the saturation altitude $z_b$, meaning no atmospheric turbulence (mixing) is triggered by the saturation. To address this issue, an efficiency factor $S^2_c$ should be introduced into (\ref{ws2}) to slightly increase the amplitude of $\hat{w}_{j,s}$. 

In practice, \citeA{lott2012stochastic} provides a version with such an efficiency factor $S_c$ to adjust the amplitude of $\hat{w}_{j,s}$,
\begin{equation}
\hat{w}_{j,s}=\frac{\Omega^2}{\vert \vec{k} \vert N} e^{-z/2H} S_{c} \frac{k^*}{\vert \vec{k} \vert}, \quad z=z_b
\label{swlott}
\end{equation}
$S_{c}$ is the saturation parameter and $S_{c}$=1.5 for Mars PCM \cite{liu2023surface};  $k^{*} = \textbf{MAX}(\vert \vec{k}_{min} \vert, 1/ \sqrt{\Delta x \Delta y})$, a parameter to prevent a monochromatic wave from occupying one whole model grid \cite{de2014intermittency}. $\Delta x$ and $\Delta y$ are the grid intervals of the model \cite{lott2012stochastic}. 

To avoid evaluating the saturation altitude $z_b$ directly, we apply the level-to-level WKB approximation of (\ref{deuxWKB}) to $\hat{w}_j$,
\begin{equation}
 \hat{w}_{j}(z_{ll+1}) =\textbf{MIN} \bigg\lbrace \hat{w}_{j}(z_{ll}) \sqrt{\frac{m_r(z_{ll})}{m_r(z_{ll+1})}} \exp \bigg ( - \int^{z_{ll+1}}_{z_{ll}} m^{ave}_i d \zeta \bigg ) , \hat{w}_{j,s} \bigg\rbrace, \quad \zeta \in [surface,top]
\label{deuxWKBs}
\end{equation}
Note that in equation (\ref{swlott}), the vertical index is $z$ instead of $z_b$. The advantage of this approach is that it allows for the use of the minimum function \textbf{MIN} in (\ref{deuxWKBs}) to automatically evaluate the vertical evolution of $\hat{w}_j$, rather than specifically imposing $\hat{w}_{j,s}$ at $z_b$, which is difficult to evaluate directly. Additionally, in equation (\ref{deuxWKBs}), the imaginary part $im^{ave}_r$ inside the integral has been omitted compared to (\ref{deuxWKB}).

However, we find a way to evaluate $z_b$ or even $A$ as inspired by (\ref{deuxWKBs}). It is useful to determine the value of $\hat{w}_j(z)$ for all $\zeta \in [z_{surface}, z_{top}]$ using (\ref{deuxWKBs}). Then, define $\hat{\Pi}^2_s = \vert \max(\hat{w}_j(z)) \vert^2 N^2 \vert \vec{k} \vert^2 / \Omega^4$ and insert it into (\ref{ws2}), yielding,
\begin{equation}
z_b=\log \bigg ( \frac{1}{\hat{\Pi}^2_s}\bigg)^H
\label{zb}
\end{equation}
By inserting $\hat{\Pi}^2_s$ into (\ref{ampliA}) and using equation (\ref{imagem}), we can determine the value of $A_j$ for a monochromatic wave,
\begin{eqnarray}
    A_j&=&\vert m_r\vert^{-\frac{3}{2}} N \vert \vec{k}\vert \vert \hat{\Pi}_s\vert \exp \bigg (\int^{z_b}_0 m_i d\zeta \bigg ) \nonumber \\
    &=& \frac{z_b}{2H}N \vert \vec{k}\vert \vert \hat{\Pi}_s\vert \nonumber \\
    &=& \frac{N \vert \vec{k}\vert}{2H}  \vert \hat{\Pi}_s\vert \log \bigg ( \frac{1}{\hat{\Pi}^2_s}\bigg)^H
    \label{aj}
\end{eqnarray}
Equations (\ref{zb}) and (\ref{aj}) may provide more accurate estimates for $z_b$ and $A$ compared to other formulas previously proposed. However, these calculations must be performed after obtaining the vertical profiles of each $\hat{w}_j$. Additionally, it is clear that both $z_b$ and $A_j$ are influenced by the wave properties as well as the background atmospheric parameters, such as $N^2$ and $\Omega^4$.

\subsubsection{Critical Layer and Viscosity}
The critical layer for a gravity wave is the level where the wave's intrinsic frequency with Doppler shift $\Omega$ changes sign,
\begin{equation}
\Theta(\Omega (z_{ll+1})\times \Omega (z_{ll})) =
\left\{\begin{array}{rcl}
   0 & \mbox{for} & \Omega (z_{ll+1})\times \Omega (z_{ll})<0 \\
   1 & \mbox{for} & \Omega (z_{ll+1})\times \Omega (z_{ll})\geq 0   
\end{array}\right. 
\label{freq}
\end{equation}
The layer forces the wave to release momentum to the mean flow by altering the vertical gradient of the $\hat{w}_j$. The $\Omega $ is dominated by background wind $\bar{u}$ for a given wave with horizontal wavenumber $\vec{k}$ and phase velocity $c$. Thus, these layers have a strong "filtering" effect on the wave \cite{Roeten2022impacts,liu2023surface,liu2025diurnal}. The critical layer is applied by multiplying (\ref{freq}) to (\ref{deuxWKBs}),
\begin{equation}
 \hat{w}_{j}(z_{ll+1}) =\Theta[\Omega (z_{ll+1})\times \Omega (z_{ll})] \textbf{MIN} \bigg\lbrace \hat{w}_{j}(z_{ll}) \sqrt{\frac{m_r(z_{ll})}{m_r(z_{ll+1})}} exp \bigg ( - \int^{z_{ll+1}}_{z_{ll}} m^{ave}_i d \zeta \bigg ) , \hat{w}_{j,s} \bigg\rbrace
\label{deuxWKBsc}
\end{equation}
The advantage of this critical layer setting is to force the monochromatic wave's amplitude to zero at the critical layer without interrupting the propagation of a wave ensemble (which includes several monochromatic waves). The induced ensemble $\hat{w}$ damps a bit at each critical layer and keeps propagating to higher altitudes. 

As the wave goes upwards, the atmospheric viscosity serves as a key factor that damps the $\hat{w}$ \cite{fritts1984gravity,lott2012stochastic}. We assume the kinematic viscosity $\nu$ increases with altitude due to the decrease of the atmospheric mass density such as $\nu=\mu/\rho$ \cite{lott2012stochastic}.  Additionally, \citeA{lott2012stochastic} takes $m^{ave}_i \approx D_{eddy}^j m^3_r/\Omega$ and absorbs $D_{eddy}^j$ into the dynamical viscosity $\mu$, making the $\mu^*=\mu D_{eddy}^j$ a tunable parameter (such as $\mu^* \approx$ 0.07 in Mars PCM). It returns,
\begin{equation}
 \hat{w}_{j}(z_{ll+1}) =\underbrace{\Theta[\Omega (z_{ll+1})\Omega (z_{ll})]}_{\textbf{CL}}\textbf{MIN}\bigg\lbrace \hat{w}_{j}(z_{ll}) \sqrt{\frac{m(z_{ll})}{m(z_{ll+1})}} \exp \bigg ( {- \underbrace{\frac{\mu^*}{\rho}}_{\textbf{VD}} \overbrace{\frac{m^3_r }{\Omega}}^{\textbf{TD}} \delta z} \bigg ), \underbrace{\hat{w}_{j,s}}_{\textbf{ST}} \bigg\rbrace
 \label{WKBFULL}
\end{equation}
Equation (\ref{WKBFULL}) represents the non-orographic GWs induced amplitude of the vertical velocity $\hat{w}_j$ in a complete form. The critical layers (\textbf{CL}), saturation (\textbf{ST}), kinematic viscosity (\textbf{VD}) , and (GW-induced) turbulence damping (\textbf{TD}) are considered together in terms of WKB approximation of $\hat{w}_j$. The variables of this equation are all known parameters of a given wave or background atmosphere. The term \textbf{TD} is detailed in section 3.1.2. It is convenient to apply (\ref{WKBFULL}) into (\ref{fullEPF}) to evaluate the vertical evolution of the wave's EP-flux.

\subsection{Evolution of the EP-flux}
We launch the waves at the average top of the planetary boundary layer. 
Inserting the complete WKB approximation of $\hat{w}_j$ (\ref{WKBFULL}) into the EP-flux definition of (\ref{fullEPF}) and letting the EP-flux evolution in the direction of $\vec{k}\Omega/\vert \vec{k}\Omega \vert$, yields, 
\begin{equation}
\vec{E}_j^{z_{ll+1}}=\frac{\vec{k} \Omega}{\vert \vec{k} \vert \vert \Omega \vert} \Theta[\Omega (z_{ll+1})\Omega (z_{ll})] \textbf{MIN} \Bigg\lbrace \vert \vec{E}_j^{z_{ll}} \vert e^{-2 \frac{\mu^* m^3}{\rho \Omega}\delta z} , \rho_r S_{c}^2 e^{-\frac{z_{ave}}{H}} \frac{\vert \Omega \vert^{3} k^{*2}}{2 N \vert \vec{k}\vert^4} \Bigg\rbrace
\label{finalEPF}
\end{equation}
We update the EP-flux by iteration loop from layer $z_{ll}$ to $z_{ll+1}$, in which $\delta z =z_{ll+1} - z_{ll}$ and $z_{ave} = (z_{ll} +z_{ll+1})/2$, respectively. The EP-flux at the reference altitude $\vec{E}^{z_r}_j$ is sampled from a normal distribution with a maximum truncation value of 5$\times$10$^{-4}$ kg m$^{-1}$ s$^{-2}$ \cite{liu2023surface}. 

\subsection{Divergence of the EP-flux}
The drag caused by the waves is given by $ \partial \vec{u}/\partial t = -\rho^{-1}  d \vec{E} /{dz}$. The drag of a wave ensemble with $M=$8 monochromatic waves is evaluated by the average drag of all waves. We use first-order Auto Regression (AR1) to average the drags on the winds and update the winds' tendencies between physical time steps $t$ and $t+\delta t$ by \cite{lott2012stochastic,lott2013stochastic}:
\begin{equation}
\Bigg ( \frac{\partial \vec{u}}{\partial t} \Bigg )_{GW}^{t+\delta t} =\frac{\delta t }{\Delta t} \frac{1}{M} \sum_{j=1}^{M} \frac{1}{\rho} \frac{d \vec{E}_j}{dz} + \frac{\Delta t - \delta t}{\Delta t} \Bigg ( \frac{\partial \vec{u}}{\partial t} \Bigg )_{GW}^{t}
\label{divergens}
\end{equation}
Thus we have  $C_j^2 = \Big (  \frac{\Delta t - \delta t}{\Delta t}   \Big )^p \frac{\delta t}{M \Delta t}$ , where $p=\lfloor (j-1) /M\rfloor$ is an integer. $\Delta t$ is the life cycle of the non-orographic gravity waves, which is approximately 24 h or 1 sol \cite{lott2012stochastic}. Note that the gravity waves' entire spectrum is accumulated over the typical life cycle of the waves $\Delta t$, which means that about $M\Delta t/\delta t$ harmonics are ejected, a number that easily is on the order of $10^3$. This excellent spectral resolution at a small numerical cost ($M=8$ waves are quite fast to evaluate each $\delta t$) is one of the benefits of the scheme. It is worth noticing that there is no temperature tendency to be implemented in the scheme. We expect that the temperature changes can be captured by the Mars PCM energy conservation design.

\section{Turbulence Induced by the Non-orographic Gravity Waves}
This section derives the eddy diffusion coefficient. It represents turbulence triggered at different altitudes. The method follows the NSP, linear damping assumption, and empirical analysis.
We derive the eddy diffusivity $D_{eddy}$ proposed by \citeA{lindzen1981turbulence}. It is valid primarily at the saturation altitude $z_b$ for a given monochromatic wave, as indicated by NSP-(d). The formulations by \citeA{holton1982role} and \citeA{weinstock1982nonlinear} may apply above $z_b$, consistent with NSP-(a) and NSP-(c). Below $z_b$, we adopt an exponential decay of eddy diffusivity based on retrievals from tracer's number density $n_q$, assuming 'symmetric turbulence' around $z_b$ (NSP-(b)), where $D_{eddy} \propto 1/\sqrt{n_q}$ \cite{rodrigo1990estimates}.
The turbulent flux from each wave ensemble is averaged over M = 8 monochromatic waves and incorporated into the model tendency via an AR-1 process.

\subsection{Eddy Diffusion Coefficient}
\subsubsection{Linear Damping Assumption}
The effect of GW-induced eddy diffusion is approximated by linear damping at and above the saturation altitude \cite{lindzen1981turbulence,holton1982role}. Assuming that diffusion contributes as a dissipation term in the equations for energy and  momentum, we have,
\begin{equation}
    D_{eddy}^j \frac{\partial }{\partial z^2} 
    \left\{ \begin{array}{c}
       \hat{\bar{u}}   \\
       \delta \hat{T}   
    \end{array}\right\} 
    = -m_r^2 D_{thermal}^j
    \left\{\begin{array}{c}
       \hat{\bar{u}}    \\
       \delta \hat{T}  
    \end{array}\right\}
     ,\quad z\geq z_b
    \label{lineardamping}
\end{equation}
where we assume the Prandtl number $Pr=D_{eddy}^j/D_{thermal}^j=1$. Although this assumption may be too small compared to 4 or 6 as suggested by \citeA{strobel1985energy} and \citeA{barnes1996transport}, the constant can be treated as a tunable parameter in practical implementation \cite{fritts2003gravity}. 

Equation (\ref{lineardamping}) implies that a quantity of $m^2_r D_{eddy}^j$ has contributed to the system. Consequently, the thermodynamic energy equation of (\ref{thee}) for $j$th monochromatic harmonic becomes, 
\begin{equation}
    [i\vec{k}(c-\bar{u}) + m_r^2 D_{eddy}^j]\delta \hat{T} = \hat{w}_j e^{z/2H} \Gamma , \quad z\geq z_b
    \label{theei}
\end{equation}
The term $m_r^2 D_{eddy}^j$ must be eliminated in (\ref{theei}) by turbulence (NSP-(c)) to keep the same form as (\ref{thee}). 
 This is equivalent to introducing an imaginary part of wave phase velocity, $c_i=\Im(c)$, to balance the term $m_r^2 D_{eddy}^j$ \cite{holton1982role}. Hence, corresponding to $m=m_r + i m_i$, we rewrite $c=c_r + i c_i$. Inserting this $c$ into (\ref{theei}), we know that the extra heating rate contributed by the wave is balanced by the turbulence using $c_i$,
\begin{equation}
    \vec{k}c_i \delta \hat{T}= m_r^2 D_{eddy}^j \delta \hat{T}, z\geq z_b
    \label{ci}
\end{equation}
It is convenient to use (\ref{ci}) to evaluate the $D_{eddy}^j$, i.e., $D_{eddy}^j=\vec{k} c_i/m^{2}_r$. However, the value of $c_i$ is unknown. Therefore, we need to find the relationship between $c_i$ and $m_i$ and express $D_{eddy}^j$ in terms of $m_i$. Either $m_i$ or $c_i$ represents the turbulence, and we may refer to $c_i$ as the phase velocity of turbulence.

\subsubsection{$m_i$ and $c_i$}
Assuming that the vertical wavenumber $m$ has an overall form similar to its real part as shown in (\ref{mreal}). The connection between $m$ and $c$ is through the parameter $\Omega=\vec{k}(\frac{\vec{k}}{|\vec{k}|}c-\vec{\bar{u}})$, the wave's intrinsic frequency with Doppler shift.  \citeA{lindzen1981turbulence} proposed to apply $c=c_r+ic_i$ ($i^2=-1$) to $m$ to derive the $m_i$ by using the relation provided in (\ref{ci}). It follows,
\begin{eqnarray}
  m & =& \frac{N |\vec{k}|}{\vec{k}(\vec{\bar{u}}-\vec{k}/|\vec{k}|c)} \nonumber\\
   &=& \frac{N |\vec{k}|}{\vec{k}(\vec{\bar{u}}-\vec{k}/|\vec{k}| c_r) -i m^2_r D_{eddy}^j} \nonumber\\
   &=& -\frac{N |\vec{k}| \Omega}{\Omega^2 + m^4_r D^2_{eddy}}+i\frac{N |\vec{k}|  m^2_r D_{eddy}^j}{\Omega^2 +m^4_r D^2_{eddy}}, \quad z\geq z_b
   \label{mc}
\end{eqnarray}
where the term $m^4_r (D^{j}_{eddy})^2$ in (\ref{mc}) is the turbulence contribution to the vertical wavenumber $m$. Additionally, the typical vertical wavelength of gravity waves is 15 km ($m_r=2\pi/15 \times 10^{-3} \approx$ 5$\times$10$^{-4}$ m$^{-1}$ ) and the $D_{eddy}^j$ is at 10$^6$ cm$^2$ s$^{-1}$, which makes the term $m^4_r D^2_{eddy} \approx$ 10$^{-10}$  s$^{-2}$  $ \ll \Omega^2$. This implies the term $m^4_r D^2_{eddy}$ is negligible. 
Unsurprisingly, now the first term of (\ref{mc}) is $-N\vert\vec{k}\vert/\Omega$, which is the exact definition of $m_r$ in (\ref{mreal}). By the same reasoning, we can conclude that the second term (the imaginary part) equals $m_i$, reads,
\begin{equation}
    m_i = \frac{N|\vec{k}| m^2_r D_{eddy}^j}{\Omega^2},\quad z\geq z_b
    \label{mi2}
\end{equation}
Equation (\ref{mi2}) shows the efficiency of turbulence arising from the superadiabatic thin layers at $z \geq z_b$. 
Additionally, equation (\ref{mi2}) explains why \citeA{lott2012stochastic} takes $m^{ave}_i \approx m^3_r/\Omega D_{eddy}^j$ in (\ref{WKBFULL}). The linear damping may not be accurate for the waves with short vertical wavelengths less than 10 km (Figure 1 of \citeA{hodges1969eddy}), which break mostly in the lower atmosphere. Applying (\ref{mi2}) below $z_b$ needs to introduce a tunable parameter such as the one in \citeA{lott2012stochastic}.

\subsubsection{Saturated Eddy Diffusivity}
The $m_i$ must take a specific value at the altitude of $z_b$ as indicated by NSP-(d). The $m_i$ in (\ref{mi2}) equals the $m_{i,s}$ of (\ref{imagem}) at $z_b$,
\begin{equation}
    \frac{N|\vec{k}| m^2_r D_{s,eddy}^j}{\Omega^2} = \frac{3}{2} \bigg\vert \frac{1}{N} \frac{dN}{dz} - \frac{1}{\Omega}\frac{d \Omega}{dz} \bigg\vert+\frac{1}{2H}, \quad z=z_b
    \label{mimi}
\end{equation}
Note that (\ref{mimi}) holds only at the saturation altitude for $j$th wave, in which we have applied the NSP-(c) and (d) to the thermodynamic equations to limit the enlargement of the "unstable" region (where the turbulence comes from). 
Therefore, the saturated eddy diffusivity $D_{s,eddy}^j$ inferred from (\ref{mimi}) is as,
\begin{equation}
    D_{s,eddy}^j= \frac{\Omega^4}{N^3 |\vec{k}|^3} \bigg(\frac{3}{2} \bigg\vert \frac{1}{N} \frac{dN}{dz} - \frac{1}{\Omega}\frac{d \Omega}{dz} \bigg\vert+\frac{1}{2H} \bigg), \quad z=z_b,  j \in [1,M]
    \label{deddyatsat}
\end{equation}
For the same reason, equation (\ref{deddyatsat}) is only effective at the saturated altitude $z_b$ for a monochromatic wave. 
A limitation arises in \citeA{lindzen1981turbulence}, where equation (\ref{deddyatsat}) is applied throughout the atmosphere above $z_b$. This approach does not account for the fact that the $m_{i,s}$ relationship in (\ref{imagem}) represents an extremum condition as indicated by NSP-(d), which is strictly valid only for saturated monochromatic waves at $z_b$.

Analogous to the concept of saturation amplitude of the EP-flux of (\ref{swlott}) provided by \citeA{lott2012stochastic}, we propose equation (\ref{deddyatsat}) as 'saturated eddy diffusivity' by giving it a tunable parameter,
\begin{equation}
    D_{s,eddy}^j=S_{mix} \frac{\Omega^4}{N^3 |\vec{k}|^3} \bigg(\frac{3}{2} \bigg\vert \frac{1}{N} \frac{dN}{dz} - \frac{1}{\Omega}\frac{d \Omega}{dz} \bigg\vert+\frac{1}{2H} \bigg) \quad z=z_b, j\in [1,M]
    \label{sed}
\end{equation}
Where $S_{mix}$ is a constant with a magnitude of 0.1. Equation (\ref{sed}) indicates that the saturated eddy diffusivity of a given wave is proportional to $m_{s,i}$. This equation has a "symmetric" $m_{s,i}$ that includes the perturbations of the background atmosphere Brunt-Väisälä frequency, which makes the formula more suitable for application in the upper atmosphere \cite{hodges1967generation,hodges1969eddy}. 
Note that both (\ref{mi2}) and (\ref{sed}) imply $D_{eddy}^j \propto m_r^{-3} \propto \lambda_z^3$, which waves with longer wavelength $\lambda_z$ provide stronger mixing. This is introduced mainly due to the linear damping assumption.

\subsubsection{Eddy Diffusivity above $z_b$}

We expect the wave momentum to be released intensively once it achieves saturation, as mentioned before. This is true, as shown in \citeA{liu2023surface}. This implies that the induced turbulence should also hit its maximum along with the saturation (NSP-d) and evolve with the damping of the momentum at higher altitudes (NSP-a). According to NSP-a that turbulence $\propto \nabla \vec{E}_j$, it prompts us to work towards establishing a relationship between $D_{eddy}^j$ and the EP-flux $\vec{E}_j$. 

However, the eddy diffusivity of the upper atmosphere derived from observations \cite{rodrigo1990estimates} shows that there are still 10$^{5}$ to 10$^{7}$ cm$^2$ s$^{-1}$ mixing excited above the turbopause or even above the exobase. We would like to point out that this mixing should not be attributed to gravity waves, since the EP-flux decays so fast after the saturation that it could cause any observable turbulence. This suggests that the molecular diffusion dominates above the turbopause \cite{slipski2018variability}.

The original algorithm to build up the connections between $D_{eddy}^j$ and the wave's EP-flux is proposed by \citeA{holton1982role} and \citeA{weinstock1982nonlinear}. Here we follow similar formalisms, but the derivations are based on the NSP and linear damping assumptions.

We start with the EP-flux definition of (\ref{fullEPF}). By inserting the WKB solution of (\ref{WKB}) into (\ref{fullEPF}), and using the $A^2-z_b$ relation of (\ref{ampliA}), the EP-flux integral function with independent variable $z$ reads,
\begin{eqnarray}
\vec{E}^z_j &=& -\rho \frac{\vec{k}}{2 \vert \vec{k} \vert^2}  m_r  \vert \hat{w}_{j}\vert^2  \nonumber \\    
    &=&- \rho \frac{\vec{k}}{2 \vert \vec{k} \vert^2} A^2 e^{-2 \int^z_{0} m_i d \zeta } \nonumber \\
    &=&- \rho \frac{\vec{k}}{2 \vert \vec{k} \vert^2} A^2 e^{-2 \int^{z_b}_0 m_i d \zeta } e^{-2 \int^z_{z_b} m_i d \zeta } \nonumber \\
    &=&\rho \frac{\Omega^3}{2 N\vert \vec{k} \vert^2} \frac{\vec{k}}{\vert \vec{k} \vert} e^{- \int^{z_b}_0 \frac{1}{H} d \zeta } e^{-2 \int^z_{z_b} m_i d\zeta } , \quad \zeta\in [surface,top]
\label{epfluxinmi}
\end{eqnarray}   

The term $\exp-2\int^z_{z_b} m_i d\zeta$ in equation (\ref{epfluxinmi}) is contributed by turbulence damping above $z_b$. As mentioned in sections 2.4.1 to 2.4.3, we rely on $m_i$ to regulate the amplitude of the turbulence. We expect $m_i$ to do the same above the saturation altitude $z_b$.
The drag caused by the wave's momentum release above $z_b$ follows,
\begin{eqnarray}
    -\frac{1}{\rho} \frac{\partial \vec{E}^z_j }{ \partial z} &=& - \frac{\partial}{\partial z} \bigg( \frac{\Omega^3}{2 N\vert \vec{k} \vert^2}  e^{- \int^{z_b}_0 \frac{1}{H} d \zeta } e^{-2 \int^z_{z_b} m_i d \zeta } \bigg) \nonumber\\
    &=&m_i \frac{\Omega^3}{ N\vert \vec{k} \vert^2}  e^{- \int^{z_b}_0 \frac{1}{H} d \zeta } e^{-2 \int^z_{z_b} m_i d \zeta } , \quad \zeta\in[surface,top]    
    \label{drags}
\end{eqnarray}
Here $\vec{k}/|\vec{k}|$ drops since the derivation takes along $z$. 

We apply $m_i=D_{eddy}^jm_r^3/\Omega$ for the $m_i$ in (\ref{drags}) to consider the evolution of the divergence of the EP-flux above $z_b$.  It follows,
\begin{equation}
 D_{eddy}^j=\frac{\Omega}{N^2\vert \vec{k}\vert} \bigg \lbrace -\frac{1}{\rho} \frac{\partial \vec{E^z_j} }{ \partial z} \bigg \rbrace  e^{\int^{z_b}_0 \frac{1}{H} d \zeta } e^{2 \int^z_{z_b} m_{i} d \zeta }, \quad z\geq z_b
 \label{deddyzgzb}
\end{equation}
Here we have applied the linear damping assumption, therefore,(\ref{deddyzgzb}) holds at $z\geq z_b$.
The last two exponential terms of (\ref{deddyzgzb}) should be considered as a constant for convenience. 
we shape the terms into a tunable parameter called $\alpha_{eff}$,
\begin{eqnarray}
 \alpha_{eff}&=&e^{\int^{z_b}_0 \frac{1}{H} d \zeta } e^{2 \int^z_{z_b} m_{i} d \zeta } \nonumber \\
 &=&\exp \bigg(-\frac{z}{H}+3\bigg\vert\frac{1}{N} \frac{\partial N}{dz} - \frac{1}{\Omega} \frac{\partial \Omega}{dz} \bigg\vert_{z_b}^z \bigg)  \nonumber \\
&\approx& \exp(-\frac{z-z_b}{H}), \quad z\geq z_b
       \label{aeff}
\end{eqnarray}
The wave's depletion altitudes are always 20 to 40 km above its saturation altitude. Here the term $\exp[-(z-z_b)/H]$ forms a factor $\alpha_{eff}$  between 0.01 ($e^{-4}$) to 0.2 ($e^{-2}$). The factor indicates the distance effect between $z_b$ and the altitudes above. 
It is worth noticing that we partially applied the $m_{i,s}$ at $z_b$ directly into the altitudes above. This bold action was first made by \citeA{holton1982role}. Our best explanation for this is that we consider the variation of $m_i$ in altitudes to be linear since it is a slowly changing parameter. The influence has been absorbed by the term $\alpha_{eff}$, 
\begin{equation}
D_{eddy}^j=\alpha_{eff}\frac{\Omega}{N^2\vert \vec{k}\vert} \bigg \lbrace -\frac{1}{\rho} \frac{\partial \vec{E^z_j} }{ \partial z} \bigg \rbrace, \qquad z\geq z_b
\label{edff}
\end{equation}
In addition, (\ref{edff}) can be achieved by inserting $m_{i,s}$ into (\ref{deddyzgzb}) in both terms. This implies that the NSP-d constraint $|\Re(d \delta \hat{T}/dt)|=\Gamma$ and the linear damping assumption in (\ref{lineardamping}) share similarities at altitudes $z\geq z_b$.

Since "saturation" and "linear damping" have been introduced to the derivation, which leads $D_{eddy}^j \propto m_r^{-3} \propto \lambda_z^3$. The relation in (\ref{edff}) favors the middle-upper atmosphere since waves with longer wavelengths break at higher altitudes. Meanwhile, the eddy diffusivity will be underestimated at the lower atmosphere where the breaking of short wavelength waves takes place. Therefore,
(\ref{edff}) is suitable for levels above the wave's saturation altitude.

\subsubsection{Eddy diffusivity below $z_b$}
The assumption of turbulence triggered by wave saturation would lack the capability to predict mixing below $z_b$. Therefore, we have to change the concepts of the NSP as mentioned before. Here we focus on the NSP-a and NSP-b.
In other words, we assume if there is momentum released from the wave, there is turbulence generated; otherwise, no turbulence. This assumption is generally true since we have observed wave-induced eddy diffusivity both below and above the saturation altitude from observations \cite{mcelroy1972stability,nier1976structure,anderson1978mariner,shimazaki1989photochemical,rosenqvist1995reexamination,sprague2004mars,slipski2018variability,yoshida2022variations}.

Now, the problem is how to evaluate the eddy diffusion coefficient below the wave's breaking altitude. It has been confirmed by many observations that the coefficient follows a power law to the long-existed tracers' abundances below the upper atmosphere \cite{von1980upper,yoshida2022variations},
\begin{equation}
D_{eddy}^j=\mathcal{B} \times n^{-1/2}_q
\label{pwl}
\end{equation}
where $\mathcal{B}$ is a constant of magnitude of 10$^{13}$ to 10$^{14}$ \cite{rodrigo1990estimates,yoshida2022variations}; $n_q$ is the number density (in cm$^{-3}$) of the tracer $q$. We always assume the density $n\propto \exp(-\frac{z}{H})$. Equation (\ref{pwl}) indicates a monotonically growing eddy diffusion coefficient that is proportional to $e^{z/H}$ below $z_b$. It equals the saturated eddy diffusivity at $z_b$. It is not very hard to guess that equation (\ref{pwl}) is equivalent to \cite{holton1982role,holton1983influence,garcia1985effect},
\begin{equation}
    D_{eddy}^j=D_{eddy}^j(z_b) e^{\beta_{diff}(z-z_b)/H},\qquad z < z_b
    \label{pwlfinal}
\end{equation}
where $\beta_{diff}$ is a positive constant that represents the growth rate of the diffusion coefficient. It turns out that $\beta_{diff}=1.5$ \cite{garcia1985effect}. This equation is a purely empirical relationship derived from observations \cite{yoshida2022variations}. However, it fully satisfies NSP-a and NSP-b. The drag in below does decrease exponentially from its peaks at saturation level \cite{liu2023surface}. In other words, equation (\ref{pwlfinal}) reveals the same nature as (\ref{edff}): the turbulence triggered by the wave is proportional to the wave's divergence of the momentum (NSP-a).

\subsubsection{A Comprehensive Formula}
A comprehensive formula of the eddy diffusion coefficient $D_{eddy}^j$ of $j$th monochromatic harmonic for the whole atmosphere can be given by combining (\ref{sed}), (\ref{edff}), and (\ref{pwlfinal}),
\begin{equation}
D^j_{eddy}= \left\{\begin{array}{rcl} \textbf{MIN} \bigg[ \alpha_{eff}  \frac{\Omega}{N^2|\vec{k}| } \bigg ( -\frac{1}{\rho} \frac{\partial \vec{E}}{\partial z} \bigg ), \quad   S_{mix}\frac{\Omega^4}{N^3 |\vec{k}|^3} \bigg(\frac{3}{2} \bigg\vert \frac{1}{N} \frac{dN}{dz} - \frac{1}{\Omega}\frac{d \Omega}{dz} \bigg\vert+\frac{1}{2H} \bigg)     \bigg] & \mbox{for} &   z \geq z_b  \\

\ D_{eddy}^j(z_b) \exp\bigg[ \beta_{diff} \frac{(z-z_b)}{H} \bigg]& \mbox{for} &  z < z_b \end{array}\right .
\label{deddy}
\end{equation}
Here we have three tunable parameters: the effective mixing factor $\alpha_{eff} =0.1 $ \cite{imamura2016convective}; the saturated factor of mixing $S_{mix}=0.1$; and diffusion decrease rate $\beta_{diff}=1.5$ \cite{garcia1985effect}. Equation (\ref{deddy}) continues at altitude $z_b$ with eddy diffusivity $D_{eddy}^j(z_b)$. The comprehensive formula makes sure the wave-induced mixing reaches its maximum at $z_b$ (NSP-d) and decreases with the divergence of the wave's momentum above (NSP-a). The mixing decreases from  $D_{eddy}^j(z_b)$ at altitudes below (NSP-a). In addition, it is convenient to evaluate the divergence of the wave's momentum above $z_b$ since the EP-flux has been treated in the non-orographic GWs schemes as shown in (\ref{WKBFULL}).

\subsection{Diffusion Flux}
Assuming that the diffusion occurs along the "propagation path" (i.e., in the direction of $\frac{\vec{k}\Omega}{\vert \vec{k} \Omega \vert}$ ) of the wave, the diffusion flux $\Phi^{df\bar{u}}_j$ for the mean zonal flow $\bar{u}$ associated with a monochromatic wave $j$ thus can be expressed as,
\begin{equation}
    \Phi^{df\bar{u}}_j =\rho \frac{\vec{k} \Omega}{|\vec{k}||\Omega|} D^{j}_{eddy} \frac{\partial \bar{u}}{\partial z}
    \label{udifflux}
\end{equation}
Note that only the zonal diffusion flux has been implemented, corresponding to the non-orographic gravity wave's EP-flux \cite{lott2012stochastic,lott2013stochastic,liu2023surface}. Here the minus sign in front $\rho$ is omitted, as the corresponding tendency equation already includes a minus sign in front of $1/\rho$; $\rho$ the background atmospheric mass density. The operator $\rho D^j_{eddy} \partial/\partial z$ turns the diffusion into momentum flux for a given field such as zonal wind.

This diffusion process also results in a mixing flux for the potential temperature $\theta$, given by
\begin{equation}
    \Phi^{df\theta}_j =\rho \frac{\vec{k} \Omega}{|\vec{k}||\Omega|} D^{j}_{eddy} \frac{\partial \theta}{\partial z}
    \label{ptflux}
\end{equation}
We expect that the atmosphere parcels involved in this mixing rise or descend adiabatically, without including significant instability in the potential temperature profile. Thus, the superadiabatic region where the turbulence comes from is stabilized and will not expand further by applying (\ref{ptflux}).

The tracers are assumed to be mixed by the turbulence. The mixing flux for a given tracer $q_k$ is similarly described by,
\begin{equation}
    \Phi^{dfq_k}_j =\rho \frac{\vec{k} \Omega}{|\vec{k}||\Omega|} D^{j}_{eddy} \frac{\partial q_k}{\partial z}
    \label{fluxqk}
\end{equation}
Here the term $\rho \frac{\vec{k} \Omega}{|\vec{k}||\Omega|} \bigg(\frac{q_k}{T}\frac{\partial T}{\partial z} +\frac{q_k}{H} \bigg)$ \cite{rodrigo1990estimates} is neglected due to its value being significantly smaller than that of $\Phi^{dfq_k}_j$ in equation (\ref{fluxqk}). More complex mixing implementation, taking into account factors such as the density gradient of tracers, especially for non-Lagrangian tracers, will be considered in a dedicated paper. In total, 43 tracers are mixed by this scheme within the Mars PCM model. The turbulence is expected to change the tracers' concentrations instantaneously. The impacts on the chemical cycle of the atmosphere will be handled by advanced chemical modules in Mars PCM.

\subsection{Turbulence Mixing}
The diffusion flux $\Phi^{df\bar{u}}_j$ can trigger zonal wind tendency since (\ref{udifflux}) is equivalent to momentum flux as mentioned before. The tendency is added to the mean flows using an AR-1 algorithm similar to Equation (\ref{divergens}),
\begin{equation}
\Bigg ( \frac{\partial \vec{u}}{\partial t} \Bigg )_{mix}^{t+\delta t} =\frac{\delta t }{\Delta t} \frac{1}{M} \sum_{j=1}^{M} \frac{1}{\rho} \frac{d \Phi ^{df \bar{u}}_i} {dz} + \frac{\Delta t - \delta t}{\Delta t} \Bigg ( \frac{\partial \vec{u}}{\partial t} \Bigg )_{mix}^{t}
\label{mixdrag}
\end{equation}
Here the tendency is only implemented in the zonal flow same as Equation (\ref{divergens}). Note that this "drag" is induced by non-orographic GWs' turbulence. The "drag" in (\ref{divergens}) is triggered by non-orographic GWs. The maximum magnitudes of turbulence-caused drags are only 10\% to 20\% or even smaller than non-orographic GWs-induced drags. 

Similarly, the mixing to the potential temperature $\theta$ is,
\begin{equation}
\Bigg ( \frac{\partial \theta}{\partial t} \Bigg )_{mix}^{t+\delta t} =\frac{\delta t }{\Delta t} \frac{1}{M} \sum_{j=1}^{M} \frac{1}{\rho} \frac{d \Phi ^{df\theta}_i} {dz} + \frac{\Delta t - \delta t}{\Delta t} \Bigg ( \frac{\partial \theta}{\partial t} \Bigg )_{mix}^{t}
\end{equation}
The background atmosphere can keep "stable" stats by applying AR-1 to the mixing-induced zonal wind and potential temperature tendency.

Two types of implementation of mixing to the tracers have been tested. (i) The impacts to the tracers $q_k$ by turbulence can be applied instantaneously, yielding,
\begin{equation}
 \frac{\partial q_k}{\partial t}  = \frac{1}{M} \sum_{j=1}^{M} \frac{1}{\rho} \frac{d \Phi ^{dfq_k} _i} {dz} 
\end{equation}
Or (ii) we apply the AR-1 average over $\Delta t=1$ sol, read,
\begin{equation}
 \bigg(\frac{\partial q_k}{\partial t} \bigg)_{mix}^{t+\delta t}  =\frac{\delta t}{\Delta t} \frac{1}{M} \sum_{j=1}^{M} \frac{1}{\rho} \frac{d \Phi ^{dfq_k} _i} {dz} + \frac{\Delta t-\delta t}{\Delta t} \bigg (\frac{\partial q_k}{\partial t} \bigg)^t_{mix}
\end{equation}
The test simulations show that the two types of implementations have similar impacts on the model's temperature and densities. 
We expect the tracers to be mixed by the turbulence, i.e., bumping up and down (determined by the waves' path vector $\vec{k}\Omega/\vert \vec{k}\Omega \vert$) in a steady atmosphere. At the same time, the model's chemical-photochemical scheme and other dynamics-related modules could take care of the tracers' reactions and trajectories. 

\subsection{The Saturation Altitude}
\begin{figure}
\centering
\includegraphics[width=13 cm,trim={1cm 0cm 3 0cm}]{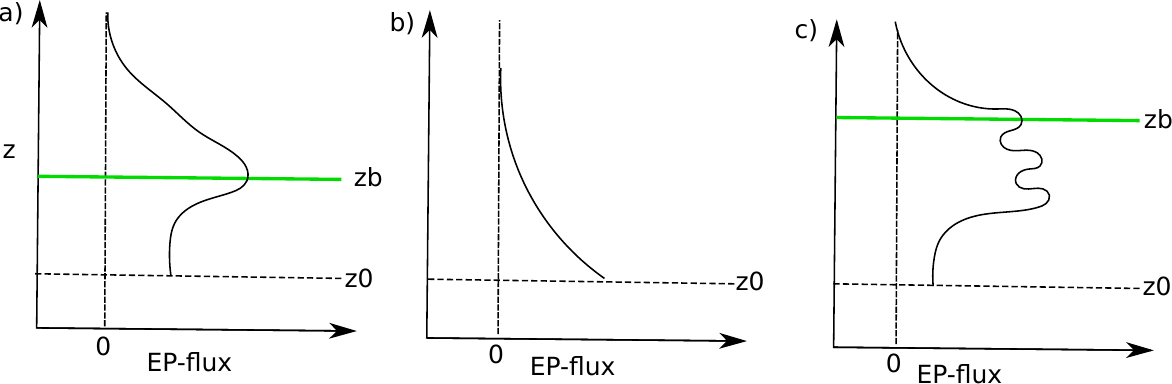} 
\caption{Wave saturation altitudes for three basic forms. a) normal case: $z_b$ at the altitudes where maximum of EP-flux take place; b) $z_b=z_0$, the wave breaks once it is launched; c) $z_b$ at the upper most saturation levels.}
\label{zbfor3}
\end{figure}

We can achieve an analytical form of wave saturation altitude $z_b$ from equation (\ref{ampliA}) as done by \citeA{lindzen1981turbulence} and \citeA{holton1982role}. However, the analytical equation is not practical because it includes the unknown constant $A^2$. Thus, we have to look into the maximum of equation (5) to search for the $z_b$. The scheme launched a wave ensemble that includes $M=8$ monochromatic waves at each time step at a given location. Each of the monochromatic waves has one of three basic EP-flux forms as shown in Figure \ref{zbfor3}.

Most of the monochromatic waves have a single peak as illustrated in Figure \ref{zbfor3}a, in which the saturation altitude is located where the EP-flux takes its maximum value. The second form of wave (Figure \ref{zbfor3}b) depletes fast above its launch level $z_0$. Thus, the saturation altitude is at the launch altitude ($z_0=z_b$). In some extremely rare cases, the harmonic breaks a few times after the saturation peak (Figure \ref{zbfor3}c). There are 2-5 levels between the first saturation (EP-flux maximum) and the final breaking level. In such a case, we take the last breaking level as the 
$z_b$ to avoid negative values (due to the derivative along $z$) when using equation (\ref{edff}) to evaluate the $D_{eddy}^j$. Therefore, the $D_{eddy}^j (z_b)$ for this type of monochromatic wave has been underestimated. Again, this effect can be balanced by the \textbf{MIN} function in (\ref{deddy}) and the tunable parameters.

\section{Sensitive Tests with Mars PCM}
The tests of the non-orographic gravity waves scheme on the Mars Planetary Climate Model have been done by \citeA{liu2023surface}. The results show that the model simulations with the scheme turned on successfully recovered the temperature/tide structures observed by the Mars Climate Sounder (MCS) below 100 km. Additionally, the high-altitude cold pockets presented in the MCS measurements \cite{heavens2022mars} are recovered by the model and explained 
directly for the first time. 
Furthermore, the simulated results are compatible with the NGIMS-derived densities in the upper atmosphere.

The impacts of non-orographic GW on the Martian atmosphere are significant during a whole Mars Year \cite{liu2023surface, liu2025diurnal}. Here we may only focus on demonstrating the differences between the clean-sky (Ls 60$^\circ$-90$^\circ$) and dusty season (Ls 240$^\circ$-270$^\circ$) for convenience. Detailed discussions for other periods may need dedicated papers.

The sensitive tests presented here compare the observations with the model simulations that include three types of modes: gravity waves off (GWoff), gravity waves on (GWon), and gravity waves mixing on (GWon+mixing; the mixing scheme is only available when the gravity wave scheme is turned on). The mixing scheme proposed above has now been adopted as one of the main physical packages in Mars PCM.

\subsection{Model Configuration}
The Mars Planetary Climate Model (formally known as LMD Mars GCM) has a grid resolution of 64x48x73 that represents an average horizontal resolution of 5.625$^\circ$ x 3.75$^\circ$ (longitude x latitude) and pseudo-altitudes from the model surface up to 250 km. The model has been coupled to an extended LMDZ physics package to simulate multiple atmospheric processes from the sub-surface layer to the exosphere. A configuration of MCD6.0-like (the most complete and advanced one) has been set up for the tests discussed in this paper \cite{liu2025datafor}. We tend not to describe what an MCD6.0-like configuration is here, since it might include a couple of dedicated papers to explain. However, we do select the Mars Years (MY29-MY36) to do the simulations. This is because there are multiple observations to compare to during MY29-MY36. Secondly, there are some realistic and high-quality measurement-retrieved scenarios (such as dust storms, water cycle, solar inputs, etc.) during these MYs. A total of $k=$ 43 tracers ($q_k$) have been used in the model.

\subsection{Results and Analysis}
\subsubsection{Model Simulated $D_{eddy}$}
\begin{figure}
\centering
\includegraphics[width=12 cm,trim={1.2cm 0.5cm 0.5cm 1cm}]{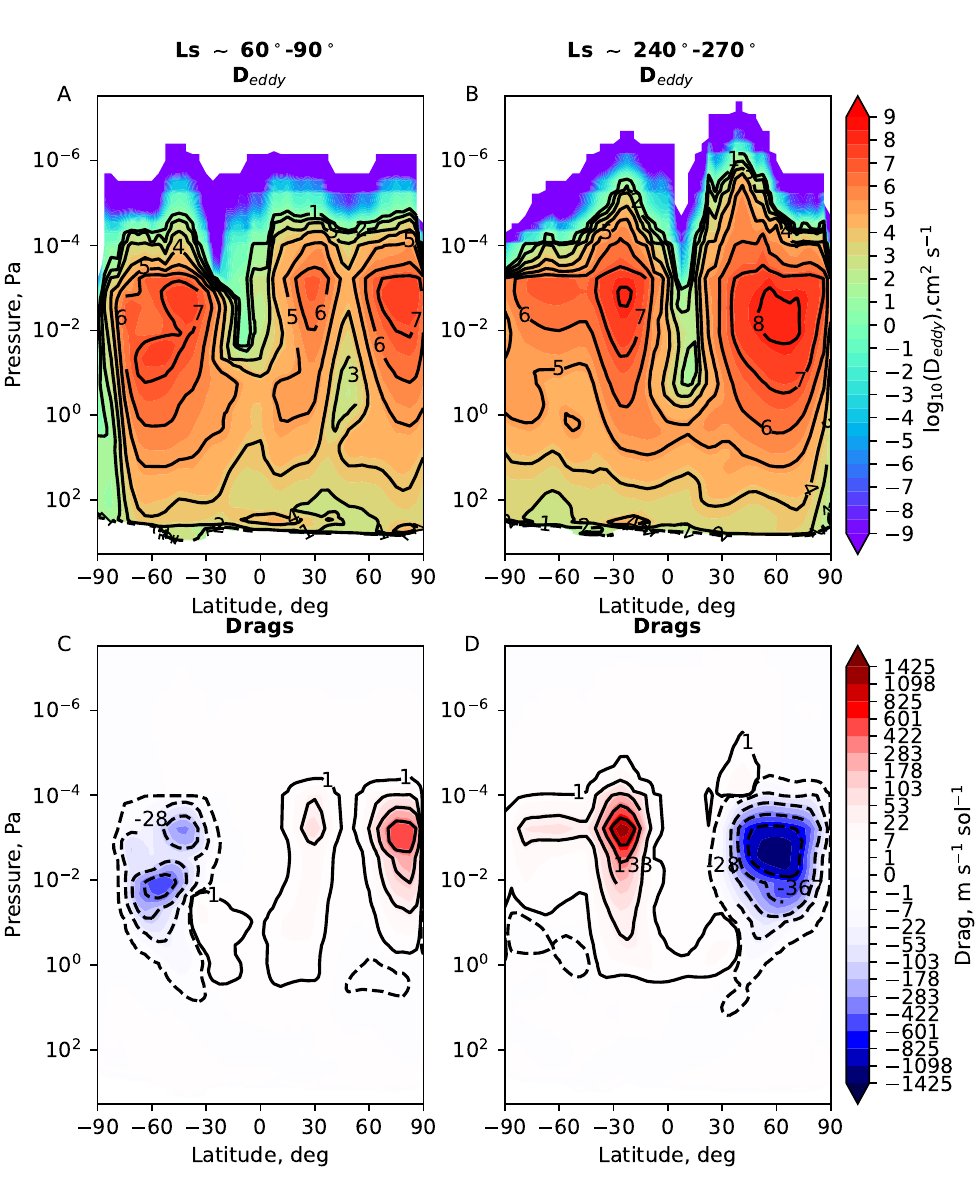} 
\caption{Monthly-averaged zonal averaged $D_{eddy}$ ( cm$^2$ s$^{-1}$, upper panels) and zonal drags (m s$^{-1}$ sol$^{-1}$, lower panels) during clear-sky (Ls 60$^\circ$-90$^\circ$ ) and dusty seasons (240$^\circ$-270$^\circ$), MY32. Note that the $D_{eddy}$ is plotted in $\log_{10}$ and the contour lines of the drags are nonlinear and not shown for values less than 10$^{0}$ cm$^2$ s$^{-1}$.}
\label{deddyetdrags}
\end{figure}

The eddy diffusion coefficients are co-located with the drags of the non-orographic drags (Figure \ref{deddyetdrags}). Three maximums exist in latitudes and in altitudes between 10$^0$ to 10$^{-4}$ Pa during the clear-sky seasons (Figure \ref{deddyetdrags}a), which correspond to the three momentum release spots (Figure \ref{deddyetdrags}c). The maximums of $D_{eddy}$ unfold as a humpbacked structure in altitudes during dusty seasons (Figure \ref{deddyetdrags}b), corresponding to the double drags peaks induced by the saturated waves (Figure \ref{deddyetdrags}d). The distribution of $D_{eddy}$ in altitude-latitude space is not a surprise since the coefficient is designed to be a ratio to the momentum release as shown in Equation (\ref{deddy}).

The coefficient lies between 10$^3$ and 10$^5$ cm$^2$ s$^{-1}$ below 10$^0$ Pa ($\approx 60 $ km) and hits maximums of 10$^6$ to 10$^7$ cm$^2$ s$^{-1}$ above during Ls 60$^\circ$-90$^\circ$(Figure \ref{deddyetdrags}a) . The diffusivity approaches zero quickly above 10$^{-4}$ Pa ($\approx $140 km) where the waves' momentum is exhausted (Figure \ref{deddyetdrags}c). There are minor mixing regions with diffusivity of 0 to 10$^3$ cm$^2$ s$^{-1}$ lying among the gaps of the drags distribution. These minima in drags are caused by the 'filtering effect' due to slow background winds at lower altitudes in the corresponding latitudes \cite{liu2023surface,liu2025diurnal}. These deep decreases and latitudinal variations in $D_{eddy}$ are consistent with derivations from NGIMS \cite{jakosky2017mars, slipski2018variability}.

The dusty seasons have larger $D_{eddy}$ of magnitudes of  10$^6$ to 10$^8$ (or even 10$^9$) cm$^2$ s$^{-1}$ above 10$^{-1}$ Pa ($\approx 80 $ km) (Figure \ref{deddyetdrags}b). A much smoother distribution of diffusivity of magnitudes of 10$^3$ and 10$^5$ cm$^2$ s$^{-1}$ than the clear-sky seasons (Figure \ref{deddyetdrags}a) lies below. Meanwhile, a deeper minor mixing region with diffusivity less than 10$^3$ cm$^2$ s$^{-1}$ appears between 6$^\circ$S and 16$^\circ$N at altitudes above 10$^0$ Pa, relating to the minimums of the waves' activities (Figure \ref{deddyetdrags}d). Additionally, the humpbacked distribution of the coefficient in the upper atmosphere makes the waves-induced turbulence extend to higher altitudes (160-180 km) during the dusty seasons than the ones (130-150 km) during clear-sky seasons.

Figure \ref{deddyetdrags} indicates a highly uneven turbopause in altitude-latitude space, which evolves seasonally. The turbopause is the boundary above which the diffusivity is dominated by molecular diffusion or the boundary where the non-orographic GWs-induced mixing is exhausted. Thus, the boundary varies between 100 and 140 km during Ls 60$^\circ$-90$^\circ$ (Figure \ref{deddyetdrags}a) and oscillates between 80 and 160 km dependent on latitudes during dusty seasons (Figure \ref{deddyetdrags}b). It is due to the enhanced intensities and saturation altitudes of the waves caused by the dust storms \cite{yiugit2015high,liu2019seasonal, Roeten2022impacts,yiugit2021dust,yiugit2021martian,liu2023surface}.

The magnitudes of the eddy diffusivity below 60 km described here are one or two orders less than the values retrieved from observations \cite{rodrigo1990estimates,anderson1978mariner,rosenqvist1995reexamination,shimazaki1989photochemical}. The orographic GWs induced mixing could explain this discrepancy since this type of wave is thought to be dominant in the lower atmosphere \cite{lott1997new,forget1999improved}. In contrast, the eddy diffusivity of the upper atmosphere is consistent with multiple observations 
\cite{sprague2004mars,yoshida2022variations}.

\subsubsection{Comparison with MCS}
\begin{figure}
\centering
\includegraphics[width=13 cm,trim={0.5cm 0.1cm 1cm 1cm}]{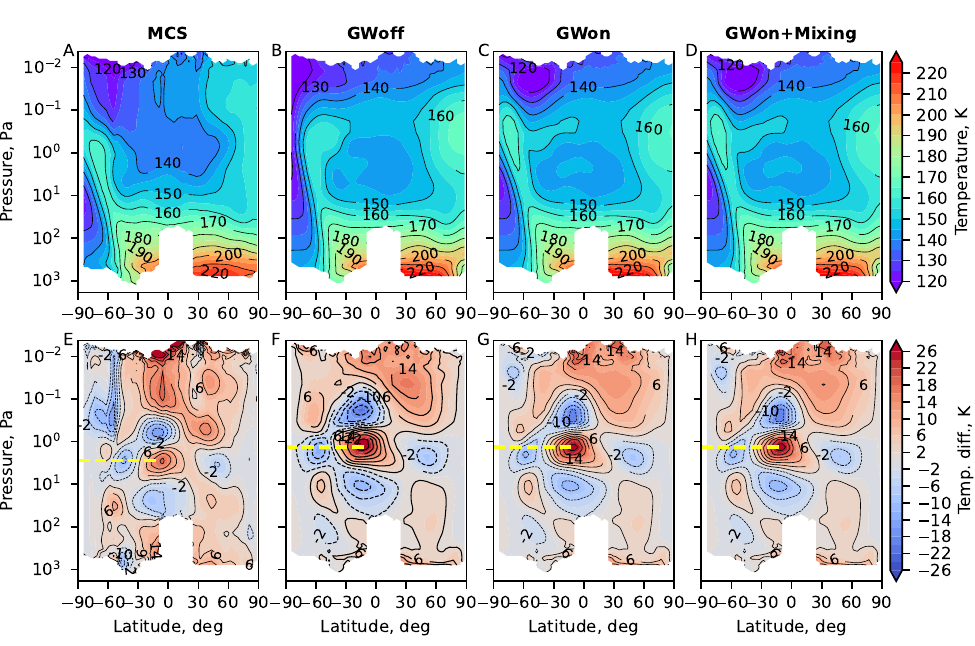} 
\caption{Monthly-averaged zonal averaged temperature (K) and diurnal tide (K), Ls 60$^\circ$-90$^\circ$, MY32. The temperature in upper panels: a) MCS observations; b) simulations with GWs, c) without GWs, and d) GWs+mixing. Lower panels e)-h) are corresponding diurnal tides.}
\label{temp03}
\end{figure}

\begin{figure}
\centering
\includegraphics[width=13 cm,trim={0.5cm 0.1cm 1cm 1cm}]{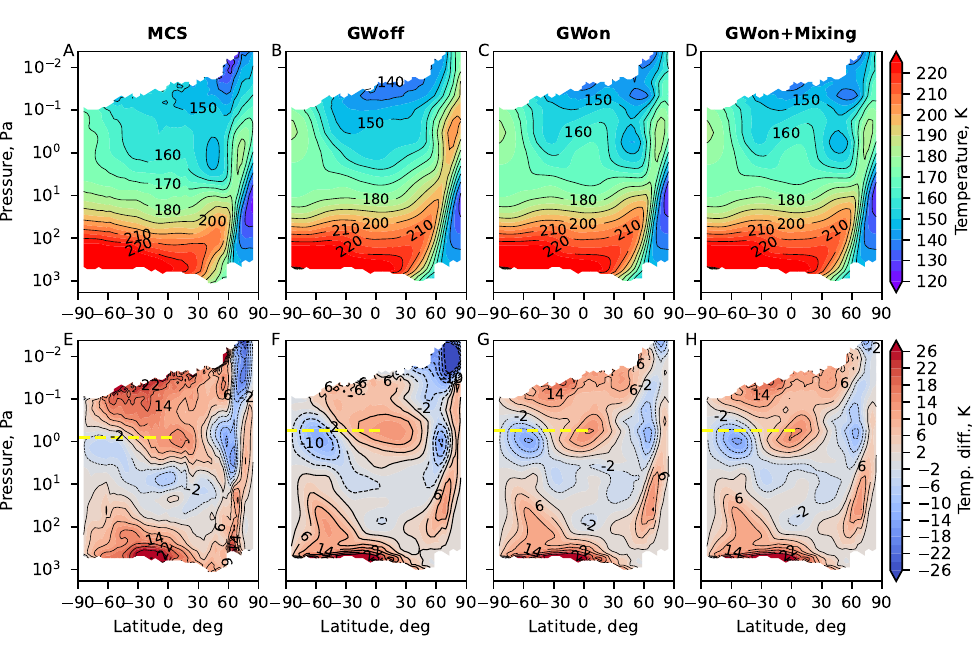} 
\caption{Similar to Figure \ref{temp03}, but for Ls 240$^\circ$-270$^\circ$, MY32.}
\label{temp09}
\end{figure}

The Mars PCM converges to the MCS temperatures below 100 km (Figure \ref{temp03}a) with turning on the GWs. A middle atmosphere warming structure lies in the simulations between 10$^0$ and 10$^{-1}$ Pa  (Figure \ref{temp03}b, \ref{temp03}c, and \ref{temp03}d). The strip may result from an inaccurate representation of the shortwave CO$_2$ heating that is represented by a wide-band scheme \cite{gonzalez2009ground}. The waves are wrongly saturated due to the structure and interact with the downward branch of the Hadley Cell. Consequently, the structure becomes even warmer in the southern hemisphere with gravity waves (Figure \ref{temp03}c and \ref{temp03}d) compared to the GWoff case (Figure \ref{temp03}b). 
Apart from that, the model predicts a high-latitude middle atmosphere "cold pocket" \cite{spiga2012gravity, heavens2022mars} with gravity waves (Figure \ref{temp03}c and \ref{temp03}d). The comparisons between simulations (Figure \ref{temp03}b, \ref{temp03}c,\ref{temp03}f, and \ref{temp03}g) and observations (Figure \ref{temp03}a and \ref{temp03}e) have been detailed in \cite{liu2023surface}. We will focus on the mixing effects in the following.

The wave-induced mixing has minor effects on the temperature (Figure \ref{temp03}d) and tide (Figure \ref{temp03}h) below 100 km (10$^{-2}$ Pa) compared to the GWon cases (Figure \ref{temp03}c and \ref{temp03}g). The mixing effects on thermal structure are also minor during the dusty seasons (Figure \ref{temp09}). This indicates that the mixing-induced drags (Equation (\ref{mixdrag})) are moderate, which is consistent with the linear damping assumption. In addition, the mixing flux is not accompanied by transferring of net heating flux , which is suggested by the Tylor-Goldstein equations and energy conservation.

\subsubsection{Impacts on Tracers}
\begin{figure}
\centering
\includegraphics[width=13 cm,trim={0.5cm 0.5cm 0cm 1cm}]{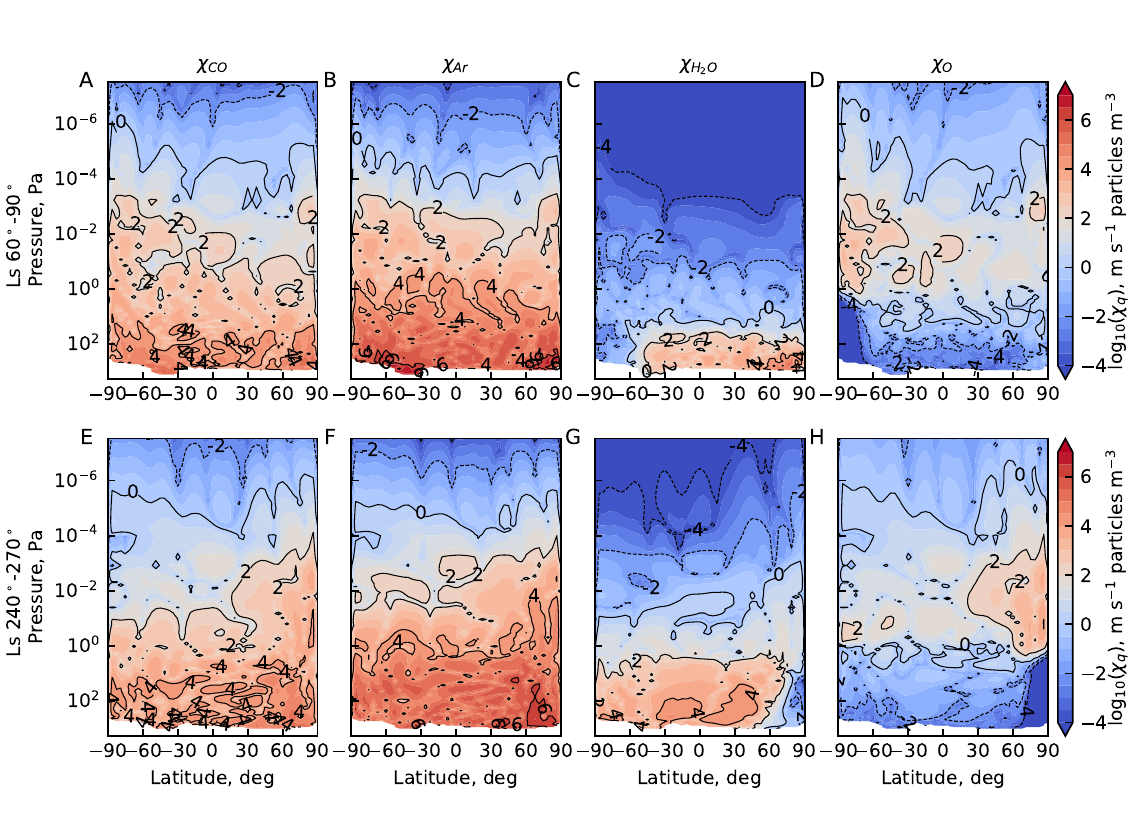} 
\caption{Monthly-averaged zonal averaged $\chi_q$ with $q$= CO, Ar, H$_2$O(vapor), O, MY32. Upper panels: $\chi_q$ during Ls 60$^\circ$-90$^\circ$. Lower panels: $\chi_q$ during Ls 240$^\circ$-270$^\circ$. }
\label{tracers}
\end{figure}

There are 43 tracers evaluated in the model. We chose to show four of them here, including CO, Ar, H$_2$O vapor, and atomic Oxygen (O). CO and Ar are considered long-existing chemical species whose lifetimes are far longer than the time scale of the eddy mixing \cite{rodrigo1990estimates,sprague2004mars,jakosky2017mars}. While O has an extremely short lifetime due to its active nature in photochemistry \cite{mahaffy2015structure}. Additionally, H$_2$O is the key to understanding Mars's habitability evolution. We define a variable $\chi_q$ to describe the mixing effects on tracers $q$, such as,
\begin{equation}
\chi_q = 10^{-6}\bigg \vert w'_{mix} n^{mix}_q -w'_{GWon} n^{GWon}_q \bigg\vert
\end{equation}
where $w'_{mix}$ and $n^{mix}_q$ are simulated vertical velocity and tracer number density with the mixing scheme turned on (GWon + mixing), respectively; $w'_{GWon}$ and $n^{GWon}_q$ are variables simulated with only GWs; the $10^{-6}$ term transfers cm$^{-3}$ into m$^{-3}$ thus $\chi_q$ has a unit of m s$^{-1}$ m$^{-3}$. Therefore, $\chi_q$ represents the differences in tracers' vertical fluxes between mixing turning on and off. Figure \ref{tracers} shows $\chi_q$ (in $\log_{10}(\chi_q)$ for convenience) in terms of CO, Ar, H$_2$O vapor, and atomic Oxygen (O).

The $\chi_{CO}$ (Figure \ref{tracers}a) experiences a fractal distribution of 10$^2$-10$^4$ below 60 km during clear-sky seasons, indicating number density variations of 10$^7$-10$^9$ cm$^{-3}$ (assuming the averaged $w'$ equals 2.5 m s$^{-1}$).
Similarly, the upper atmospheric $n_{CO}$ variations are of 10$^5$-10$^8$ cm$^{-3}$. This applies to dusty seasons as well (Figure \ref{tracers}e). A similar situation happens for Ar (Figure \ref{tracers}b and \ref{tracers}f).

A lack of middle atmosphere to lower thermosphere H$_2$O vapor exchange occurs during Ls 60$^\circ$-90$^\circ$ (Figure \ref{tracers}c) due to a relatively weaker Hadley circulation \cite{liu2023surface}, in which the density variations lie between 10$^4$-10$^6$ cm$^{-3}$. In contrast, the variations increase to 10$^5$-10$^7$ cm$^{-3}$ during dusty seasons (Figure \ref{tracers}g). Additionally, the variations of O are minor below 30 km (Figure \ref{tracers}d and \ref{tracers}h) due to the low O density in these regions. The variations are equivalent to the stable gases above 30 km.

The $\chi_q$ collocates with the asymmetric Mars Hadley Cell that is reflected by the temperatures (Figure \ref{temp03} and \ref{temp09}). It causes intensive mass exchanges in the descending branches of HC, i.e., the southern hemispheric polar region during Ls 60$^\circ$-90$^\circ$. Likewise, the intense matter (CO$_2$ condensation) exchanges happen at the northern polar area during Ls 240$^\circ$-270$^\circ$. Therefore, it is not very surprising to observe Ar enhancement (a kind of 'alien weather' as termed by \citeA{forget2004alien} ) in the observations \cite{sprague2004mars} during the clear-sky southern high-latitude region as shown in Figure \ref{tracers}b. 

The mixing on tracers is coupled with the dust activities (lower panels of Figure \ref{tracers}). The altitudes of the mixing increase by 20-40 km during dusty seasons, facilitating the transport of H$_2$O into the thermosphere much more easily (Figure \ref{tracers}g). The mixing effects are stronger during Ls 240$^\circ$-270$^\circ$. Therefore, the H$_2$O spikes in the lower thermosphere would be common during the dusty seasons, which have been noticed by observations \cite{fedorova2020stormy,stone2020hydrogen,chaffin2021martian}. The atmosphere experiences an overall expansion during the dusty seasons, in which the atmospheric densities are increased at a given altitude. This decreases the kinematic viscosity $\nu=\mu/\rho$ (Equation \ref{WKBFULL}), favoring the waves' momentum deposition at higher altitudes. On the other hand, the density increases, and the wind perturbations have strengthened the waves' EP-flux (Equation \ref{fullEPF}), leading to higher momentum intensity during the dust seasons (Figure \ref{deddyetdrags}d). This generates higher $D_{eddy}$ values (Figure \ref{deddyetdrags}b) than the eddy diffusivity during clear-sky seasons (Figure \ref{deddyetdrags}a). It causes intensive and almost instantaneous mixing and transport of tracers between the middle atmosphere and the thermosphere during the dusty seasons.

\subsubsection{Comparison with NGIMS}

\begin{figure}
\centering
\includegraphics[width=8cm,trim={1.3cm 0.3cm 1cm 0.5cm}]{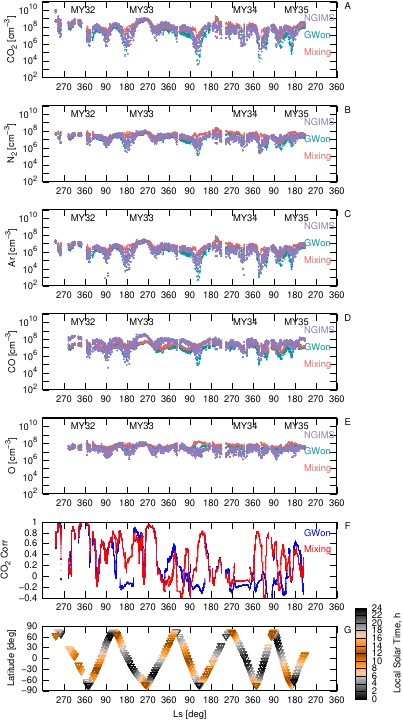} 
\caption{Neutrals abundance (cm$^{-3}$) at altitudes of 180$\pm$0.1 km,from NGIMS (purple), GWon (shallow blue), and Mixing (shallow red), MY32 to MY35. a) CO$_2$; b)N$_2$; c) Ar; d) CO; and e) O; f) correlated coefficients of a) between NGIMS and simulations (GWon in blue and Mixing in red); g) observational geometry. The duplicated observations in (d) are due to the difficulty in separating CO from N$_2$ by NGIMS.}
\label{ngimsgwonmixing}
\end{figure}

Figure \ref{ngimsgwonmixing} shows the density comparisons between NGIMS-measured five neutrals (CO$_2$, N$_2$, Ar, CO, O) and the simulations with and without mixing at altitudes of 180 km during MY32 to MY35. The correlated coefficients between the observations and simulations are shown in Figure \ref{ngimsgwonmixing}f for the CO2 case (Figure \ref{ngimsgwonmixing}a).
The observational geometry is illustrated in the bottom panel (Figure \ref{ngimsgwonmixing}g). Note that the simulations are instantaneous outputs every 2 hours. However, the observations are high-frequency samples (1 Hz) during short periods ( $\approx$ 300 s, inbound). The outputs of the simulations are interpolated linearly to compare with the observations.

A net increase in all of the five species is caused by the mixing and related chemical processes (Figure \ref{ngimsgwonmixing}). 
The large simulation-observation discrepancies in MY33, MY34, and MY35 during Ls 0$^\circ$-180$^\circ$ are compensated for by the mixing effects (Figure \ref{ngimsgwonmixing}a to \ref{ngimsgwonmixing}c). The increase or decrease of the upper densities is a complicated response of the upper atmosphere to the dynamics induced by the drags and turbulence of gravity waves.

The abundance of CO$_2$ is improved due to the turbulence of the waves (Figure \ref{ngimsgwonmixing}a), leading to a better matching with the observations during all MYs except MY34. The model has an overestimate of upper atmosphere density during the global dust event (MY34) due to over-ejected dust. The N$_2$ and Ar (Figure \ref{ngimsgwonmixing}b and \ref{ngimsgwonmixing}c) follow the similar trends as CO$_2$. NGIMS has difficulties in separating CO from N$_2$ \cite{benna2014datasets}, thus resulting in overlapped values in the observations (Figure \ref{ngimsgwonmixing}d). The simulations that include mixing show a higher overall correlation with NGIMS compared to the GWon-NGIMS correlations (Figure \ref{ngimsgwonmixing}f).

Besides, the reverse evolution in the simulations (Ls 300$^\circ$, MY33 to Ls 120$^\circ$, MY34; Ls after 200$^\circ$, MY35) may indicate an overconsumption of CO due to chemical reactions during the day (LST 06-18 h, Figure \ref{ngimsgwonmixing}g). Figure \ref{ngimsgwonmixing}e shows that the simulations lack diurnal variations in atomic oxygen. This may be due to the much longer model output time intervals compared to the extremely short lifetime of atomic oxygen. In addition, all five species are overestimated by the mixing scheme during nighttime. This suggests that the nighttime mixing needs to be weaker and the GWs need to be emitted from a lower altitude to increase the 'wind filtering' effects \cite{liu2025diurnal}.

\section{Summary}
The turbulence triggered by non-orographic Gravity Waves (GWs) is implemented in the Mars Planetary Climate  Model (Mars PCM) using stochastic parametrization. We revisit and refine the formalisms of \citeA{lindzen1981turbulence}, \citeA{holton1982role}, and \citeA{weinstock1982nonlinear} to better capture the mixing of non-orographic gravity waves. The new formalism integrates with a stochastic GW scheme designed by \citeA{lott2012stochastic}, which operates across the entire atmosphere in the Mars PCM \cite{liu2023surface}. The scheme has been tested with the Mars PCM. Simulations are compared with temperatures from the Mars Climate Sounder (MCS, onboard Mars Reconnaissance Orbiter) below 100 km, and abundances from the Neutral Gas and Ion Mass Spectrometer (NGIMS, onboard the Mars Atmosphere and Volatile EvolutioN mission) in the upper atmosphere.

A stochastic turbulence scheme for non-orographic GWs is developed and integrated into the framework of \citeA{liu2023surface}, with both the GW and turbulence formulations derived from a refined Non-Superadiabatic Principle (NSP).

The concept of NSP originally links GW-induced turbulence to thin superadiabatic layers caused by wave saturation \citeA{hodges1967generation,hodges1969eddy,lindzen1981turbulence,holton1982role}. The idea has been refined without major changes to the formulations. Under the new parameterization, turbulence is assumed to arise from wave momentum release. This allows turbulence to be generated throughout all altitudes, rather than only at the saturation level.

The refined NSP can describe wave saturation and saturated turbulence via the energy equation consistently. By applying linear damping to represent turbulent dissipation, it suggests that the \citeA{lindzen1981turbulence} formulation is suitable to represent the saturated turbulence at $z_b$. The \citeA{holton1982role} approach works above $z_b$. The empirical formula applied below $z_b$ is in accordance with the refined NSP.
Thus, the wave-induced eddy diffusivity can be given by a comprehensive formula for all atmospheric layers. The coefficient is proportional to the divergence of the GW's EP flux as suggested by the refined NSP.

The turbulent diffusion is represented by the averaged diffusion flux of a wave ensemble in zonal winds, tracers, and potential temperature. The flux divergences are then added to the mean fields using a first-order Auto-Regressive algorithm. This approach was first proposed by \citeA{lott2012stochastic} to capture the drags of a wave ensemble.

Therefore, the non-orographic GW parameterizations of \citeA{lott2012stochastic} and \citeA{liu2023surface} are theoretically integrated with the new turbulence scheme via refined NSP and linear damping in the wave energy equation.

Simulations conducted by Mars PCM show that a maximum of 10$^6$ to 10$^7$ cm$^2$ s$^{-1}$ eddy diffusion coefficient has been triggered at 10$^0$ to 10$^{-4}$ Pa due to the momentum released from the waves during the clear-sky seasons. The value hits 10$^7$ to 10$^8$ (or even 10$^9$) cm$^2$ s$^{-1}$ during dusty seasons due to the enhanced intensity of the momentum and wave-breaking altitude.

This seasonally and spatially varying eddy diffusion plays a key role in shaping the upper atmosphere, especially near the turbopause — the transition region where eddy and molecular diffusion are of similar magnitude. Mars has a turbopause that varies from 70 to 140 km, dependent on latitude and seasons, implying the key role of GW-induced turbulence in atmosphere escape \cite{jakosky2017mars, slipski2018variability}.

The turbulence has minor effects on the model temperature and tide, suggesting no net thermal energy is emitted. Therefore, the scheme conserves energy.

The estimated turbulence has a strong vertical transport effect on the tracers such as CO, Ar, H$_2$O(vapor), and O. This can explain some instantaneous
tracers' density fluctuations observed by observations \cite{stone2020hydrogen}.

Comparisons with NGIMS suggest that the turbulence from non-orographic GWs can regulate the upward/downward transport of atmospheric species.

The overestimate of nighttime upper atmospheric abundances may be caused by weak GW buildup during the night. This leads to insufficient cooling. The problem occurs when waves are launched from an averaged planetary boundary layer (PBL). A better approach is to vary the source altitude with the diurnal cycle of the Martian PBL \cite{liu2025diurnal}. More advanced mixing implementation considering the tracers' density gradients will be considered in the future.

\section*{Open Research Section}
The NGIMS density datasets \cite{benna2014datasets} and MCS temperature profiles \cite{mccleese2008datasets} analyzed in this study are publicly accessible through the NASA Planetary Data System. Simulations were conducted using the Mars Planetary Climate Model (PCM), freely available at \url{svn.lmd.jussieu.fr/Planeto/trunk}, with revision r3263. All results are reproducible using the provided files and instructions in \citeA{liu2025datafor}.

\acknowledgments
J. Liu acknowledges the National Natural Science Foundation of China (Grant No. 42241115). We also extend our gratitude to GENCI-CINES for providing HPC computing resources (Grant No. A0160110391). We acknowledge funding from the European Research Council (ERC) under the European Union’s Horizon 2020 Research and Innovation Program (Grant Agreement No. 835275, project \textit{Mars Through Time}). We acknowledge the constructive efforts of the two anonymous reviewers who took the time to review such a complicated technical paper.

%
%

\bibliography{agusample}

%
%
%
%
%

\end{document}